\documentclass[a4paper,aps,prd,preprint,nofootinbib]{revtex4-1}

\usepackage{amsmath}
\usepackage{dsfont}
\usepackage{multirow}
\usepackage[hidelinks]{hyperref}
\usepackage{cleveref} 
\usepackage{color}
\definecolor{gray}{gray}{.5}
\usepackage{graphicx}

\DeclareMathOperator{\Tr}{Tr}

\begin{document} 

\preprint{MS-TP-21-06}

\title{New constraints on radiative seesaw models from IceCube \\ and other neutrino detectors}

\author{Thede de Boer$^{1}$}
\author{Raffaela Busse$^{2}$}
\author{Alexander Kappes$^{2}$}
\author{Michael Klasen$^{1}$}\email{michael.klasen@uni-muenster.de}
\author{Sybrand Zeinstra$^{1}$}

\affiliation{$^1$Institut f\"ur Theoretische Physik, Westf\"alische Wilhelms-Universit\"at M\"unster, Wilhelm-Klemm-Stra\ss{}e 9, 48149 M\"unster, Germany}
\affiliation{$^2$Institut f\"ur Kernphysik, Westf\"alische Wilhelms-Universit\"at M\"unster, Wilhelm-Klemm-Stra\ss{}e 9, 48149 M\"unster, Germany}

\begin{abstract}
Dark matter (DM) scattering and its subsequent capture in the Sun can boost the local relic density, leading to an enhanced neutrino flux from DM annihilations that is in principle detectable at neutrino telescopes. We calculate the event rates expected for a radiative seesaw model containing both scalar triplet and singlet-doublet fermion DM candidates. In the case of scalar DM, the absence of a spin dependent scattering on nuclei results in a low capture rate in the Sun, which is reflected in an event rate of less than one per year in the current \textsc{IceCube} configuration with 86 strings. For singlet-doublet fermion DM, there is a spin dependent scattering process next to the spin independent one, which significantly boosts the event rate and thus makes indirect detection competitive with respect to the direct detection limits imposed by \text{PICO-60}. Due to a correlation between both scattering processes, the limits on the spin independent cross section set by \textsc{XENON1T} exclude also parts of the parameter space that can be probed at \textsc{IceCube}. Previously obtained limits by \textsc{ANTARES}, \textsc{IceCube} and \textsc{Super-Kamiokande} from the Sun and the Galactic Center are shown to be much weaker.
\end{abstract}

\maketitle

\section{Introduction}
\label{sec:1}

Two of the biggest current challenges in theoretical particle physics concern the nature of dark matter (DM) as well as the generation of neutrino masses, both of which cannot be convincingly explained by the Standard Model (SM) \cite{Zyla:2020zbs,Klasen:2015uma}. Radiative seesaw models, as considered in this paper, connect these two open ends. In these models, neutrino masses are generated at one-loop level through interactions with a dark sector containing a DM candidate, the most famous example being the Scotogenic Model \cite{Ma:2006km,Klasen:2013jpa,deBoer:2020yyw}. The topology of the neutrino loop as well as the particle content of the dark sector allow for many different options, which have been systematically categorized in Ref.\ \cite{Restrepo:2013aga}.

The link between neutrinos and the dark sector allows for neutrino signals produced by DM annihilations. Such annihilations in the galactic DM halo can lead to high energy neutrinos that, if abundant enough, can be observed with \textsc{IceCube} and other neutrino detectors. The characteristics of the neutrino spectrum will depend on the annihilation processes involved. In particular, direct annihilation into neutrino pairs will result in a clear, distinct line at $E_\nu \simeq m_\text{DM}$, whereas neutrinos produced from the decays of other SM particles created by annihilations result in a continuous spectrum. Direct annihilation into neutrinos has been studied in Refs.\ \cite{Arina:2015zoa,Lindner:2010rr,ElAisati:2017ppn,Farzan:2011ck}. Here, we will study the general case, where both scenarios are taken into consideration.

Monochromatic neutrinos provide a clear and distinct signal compared to a continuous spectrum. Since we also take into account the latter type of signals, we consider the case where the annihilation rate is enhanced by a local boost in the relic density. We focus on annihilations due to an increased relic density in the Sun. Other astrophysical objects that have been considered in the literature are e.g. the Earth \cite{Andreas:2009hj,Albert:2016dsy}, the Galactic Center \cite{Aartsen:2017ulx,Albert:2016emp,Aartsen:2020tdl,Abe:2020sbr}, or super massive black holes \cite{Arina:2015zoa}.

As our solar system moves through the galactic halo, DM in the form of a weakly interacting massive particle (WIMP) can scatter off nuclei in the Sun. If enough of the kinetic energy of the WIMP is transferred in these scatterings, the WIMP will be gravitationally captured by the Sun \cite{Gould:1987,Gould:1991hx}. This will lead to an accumulation of WIMPs in the Sun's core, enhancing the local relic density and leading to a boost in annihilations. The capture rate depends chiefly on the WIMP-nucleon scattering cross section, on which direct and indirect detection experiments have put stringent constraints \cite{Zyla:2020zbs,Klasen:2015uma,Aprile:2018dbl,Aprile:2019dbj,Aprile:2019xxb,Amole:2019fdf,Adrian-Martinez:2016gti,Aartsen:2016zhm,Choi:2015ara}.

For definiteness, we consider here the radiative seesaw model T1-3-B with $\alpha=0$ \cite{Fiaschi:2018rky,May:2020bpo}, following the notation in Ref.\ \cite{Restrepo:2013aga}, that can contain either a scalar or a fermionic DM candidate. In Sec.\ \ref{sec:2} we give an overview of the model and in Sec.\ \ref{sec:3} of the WIMP-nucleon scattering processes that are present. Our method to compute the corresponding neutrino flux and event rate in \textsc{IceCube} is discussed in Sec.\ \ref{sec:4}. In Sec.\ \ref{sec:5} we present the results of a numerical scan of the model parameter space. We show results for the spin independent and spin dependent WIMP-nucleon cross sections, the thermally averaged cross section in the Galactic Center and the expected event rates from annihilations in the Sun in the current \textsc{IceCube} configuration with 86 strings (\text{IC86}). These results are not only compared to limits from the direct detection experiments \textsc{XENON1T} and \textsc{PICO-60}, but also to previously obtained limits from neutrino observations in the Sun and the Galactic Center with \textsc{ANTARES}, \textsc{IceCube} and \textsc{Super-Kamiokande}. Our conclusions and an outlook are given in Sec.\ \ref{sec:6}. Analytic results for mixing in the fermion sector and the resulting interaction vertices are deferred to the Appendix.

\section{The model}
\label{sec:2}

The radiative seesaw mechanism at one loop allows for $35$ models that can simultaneously generate neutrino masses and present a suitable DM candidate \cite{Restrepo:2013aga}. What these models have in common is that the new fields are odd under a $\mathds{Z}_2$ symmetry, whereas the SM fields have an even charge. As a result there must be an even number of new fields at any interaction vertex. Thus the new particles cannot decay into SM particles, which guarantees the stability of the DM candidate. Scattering and annihilation processes of new particles are allowed. When two different dark particles are close in mass, coannihilation processes can become relevant \cite{Suematsu:2009ww,Harz:2012fz,Klasen:2013jpa,Herrmann:2014kma,Harz:2014tma,Branahl:2019yot}. Moreover, the $\mathds{Z}_2$ symmetry does not allow for any tree-level seesaw mechanism. Instead neutrino masses are generated at one-loop level. The models in Ref.\ \cite{Restrepo:2013aga} are categorized based on the different topologies of the neutrino loops.

\begin{figure}
	\centering
	\includegraphics[]{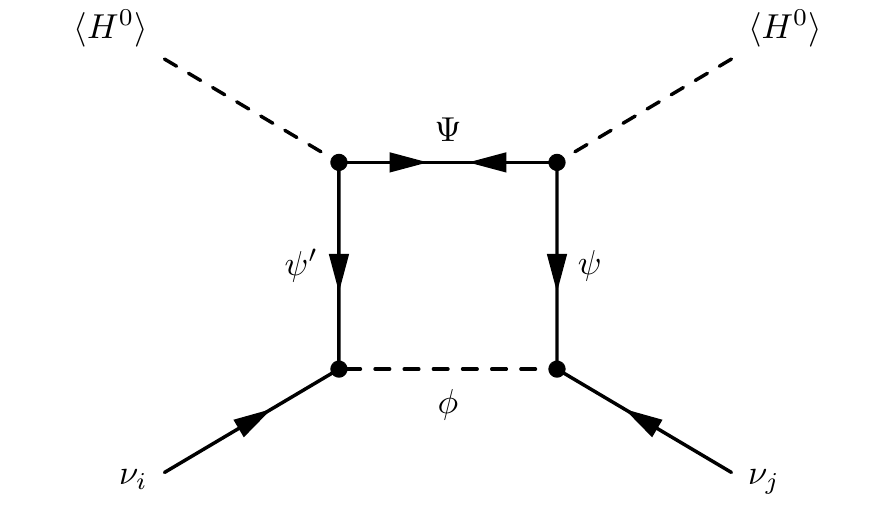}
	\caption{Topology of neutrino mass generation in the T1-3 class of radiative seesaw models \cite{Restrepo:2013aga,Fiaschi:2018rky,May:2020bpo}.}
	\label{fig. T13 loop}
\end{figure}
Figure \ref{fig. T13 loop} shows the loop of T1-3 topology. In this paper we study the model T1-3-B with $\alpha=0$.  The fields that are added to the SM in this case are listed in Tab.\ \ref{tab. NewFields}. Componentwise these fields are given by
\begin{align}
  \Psi=\Psi^0,\quad 
  \psi=
  \left(\begin{matrix}
    \psi^0\\ \psi^-
  \end{matrix}\right),\quad 
  \psi'=
  \left(\begin{matrix}
    \psi'^+\\ \psi'^0
  \end{matrix}\right), \quad 
  \phi_i=
  \left(\begin{matrix}
    \frac{1}{\sqrt{2}}\phi_i^0 & \phi_i^+\\
    \phi_i^-&-\frac{1}{\sqrt{2}}\phi_i^0\\
  \end{matrix}\right),
\end{align} 
where the index $i$ denotes the generations of the scalar triplet and the superscripts denote the electric charge. The scalar triplet has no hypercharge, therefore its neutral component is a real scalar. The two fermion doublets are combined to a vector-like doublet by building Dirac spinors as follows:
\begin{align}
  \psi^0_\text{D}=\left(\begin{matrix}
    \psi^0\\i\sigma_2(\psi'^0)^*
  \end{matrix}\right),\quad
  \psi^-_\text{D}=\left(\begin{matrix}
    \psi^-\\i\sigma_2(\psi'^+)^*
  \end{matrix}\right).
\end{align}
However, throughout this work we stick to using Weyl spinors instead of these Dirac spinors. The particle content of this model is similar to the $F_4^m$ and $S_6^r$ models considered in Ref.\ \cite{ElAisati:2017ppn}, except for the important difference that our fermions are mixed singlet-doublet states.

\begin{table}
  \centering
  \caption{New fields in the model T1-3-B with $\alpha=0$.}
  \label{tab. NewFields}
  \begin{tabular}{ccccccc}
  \hline
    Field & Type & Generations & SU$(3)_C $ & SU$(2)_L $& U$(1)_Y$ & $\mathds{Z}_2$\\
    \hline
    $\Psi$ & Majorana spinor & 1&  $1 $&$ 1$&$ 0$ & $-1$ \\
    $\psi$ & Weyl spinor & 1 & $1 $&$ 2$&$ -1$ &  $-1$ \\
    $\psi'$ & Weyl spinor & 1 & $1 $&$ 2$&$ 1$ &  $-1$ \\
    $\phi$ & Real scalar & 2 & $1 $&$ 3$&$ 0$ &  $-1$ \\
    \hline
  \end{tabular}
\end{table}

Following the notation of Refs.\ \cite{Fiaschi:2018rky,May:2020bpo}, the Lagrangian of the model is given by\footnote{The products of field multiplets must be interpreted to be gauge and Lorenz invariant.}
\begin{align}
 \mathcal{L} =&\mathcal{L}_\text{SM}+\mathcal{L}_\text{kin}
 - \frac{1}{2} (M_\phi^2)^{ij} \Tr(\phi_i \phi_j)
 - \left( \frac{1}{2} M_\Psi \Psi \Psi + \text{H. c.} \right)
 - \left( M_{\psi\psi'} \psi \psi' + \text{H. c.} \right) \nonumber\\
 &- (\lambda_1)^{ij} (H^\dagger H) \Tr(\phi_i \phi_j)
 - (\lambda_3)^{ijkm} \Tr(\phi_i \phi_j \phi_k \phi_m) \nonumber\\
 &- \left( \lambda_4 (H^\dagger \psi') \Psi + \text{H. c.} \right)
 -\left( \lambda_5 (H \psi) \Psi + \text{H. c.} \right)
 -\left( { (\lambda_6)^{ij}} L_i \phi_j \psi' + \text{H. c.} \right).
\end{align}
The $\lambda_6$ term couples the SM leptons $L_i$ to the new fields, which allows the neutrinos to obtain their masses. Since $M_\phi^2>0$, the scalar triplets does not aquire a vacuum expectation value (VEV). They couple to the SM Higgs $H$ through the $\lambda_1$ term and have self interactions through the $\lambda_3$ term. The self interactions do not influence the phenomenology of the model. Hence $\lambda_3$ is set to zero in this work. The $\lambda_4$ and $\lambda_5$ terms are similar to Yukawa terms and link the new fermions to the SM Higgs field. After electroweak symmetry breaking (EWSB) these terms will appear in the mass matrix of the fermions and induce mixing between the fermion singlet and the vector-like doublet.

To obtain the physical states (mass eigenstates), the mass matrices must be diagonalized. After EWSB, the mass matrix for the neutral fermions is given by
\begin{align}\label{eq. fermion masses}
  M_f=\left(\begin{matrix}
    M_\Psi & \frac{\lambda_5 v}{\sqrt{2}} & \frac{\lambda_4 v}{\sqrt{2}} \\
    \frac{\lambda_5 v}{\sqrt{2}} & 0& M_{\psi\psi'} \\
    \frac{\lambda_4 v}{\sqrt{2}} & M_{\psi\psi'} & 0
  \end{matrix}\right),
\end{align}
and it is diagonalized by the unitary mixing matrix $U_\chi$. This results in three Majorana mass eigenstates with masses $m_{\chi_i^0}$, given by
\begin{align} 
  \left(\begin{matrix}
    \chi_1^0\\\chi_2^0\\\chi_3^0
  \end{matrix}\right)=U_\chi
  \left(\begin{matrix}
    \Psi^0\\\psi^0\\\psi'^0
  \end{matrix}\right).
\end{align}
Due to the interaction of the scalar triplet with the SM Higgs boson, the scalar mass matrix obtains a contribution proportional to the SM Higgs VEV $v$,
\begin{align}
  M^2_{\phi^0}=M^2_{\phi^\pm}=M^2_\phi+\lambda_1v^2.
\end{align}
The scalar mass matrix is diagonalized by $O_\eta$, which yields the squared masses of the scalar components $m^2_{\eta_i^{0,\pm}}$. The mass eigenstates are defined by
\begin{align}
  \left(\begin{matrix}
    \eta_1^{0,\pm}\\\eta_2^{0,\pm}
  \end{matrix}\right)=O_\eta
  \left(\begin{matrix}
    \phi_1^{0,\pm}\\\phi_2^{0,\pm}
  \end{matrix}\right).
\end{align}
In this work we choose $M^2_\phi$ and $\lambda_1$ to be diagonal and thus neglect mixing between the two generations of scalar particles, as this does not affect the phenomenology. Note that the neutral and charged components have equal masses at tree level. Loop corrections induce a mass splitting between both components, making the charged component $166$ MeV heavier at one loop \cite{Cirelli:2005uq}.

The neutrino loop in Fig.\ \ref{fig. T13 loop} can be calculated analytically resulting in the following formula for the neutrino mass matrix:
\begin{align}
  (M_\nu)_{ij} &= \frac{1}{32\pi^2} \sum_{l=1}^{n_{\text{s}}} { \lambda_6^{im} \lambda_6^{jn}} (O_\eta)_{ln} (O_\eta)_{lm}
 \sum_{k=1}^{n_{\text{f}}} {(U_\chi)^*_{k3}}^2 \frac{m_{\chi_k^0}^3}{m_{\eta^0_l}^2 - m_{\chi_k^0}^2}  \frac{m_{\chi_k^0}^2}{m_{\eta^0_l}^2}\nonumber\\
 &=: \frac{1}{32\pi^2} \sum_{l=1}^{n_{\text{s}}} { \lambda_6^{im} \lambda_6^{jn}} (O_\eta)_{ln} (O_\eta)_{lm}\,A_l.
 \label{eq. neutrino masses}
\end{align}
Here, $n_\text{s}$ is the number of neutral scalars and $n_\text{f}$ is the number of neutral fermions. In our model with two generations of a real scalar triplet, $n_\text{s}=2$. This implies that the rank of the matrix $M_\nu\leq 2$. Thus this model only allows for two massive neutrinos. One could consider the case with three massive neutrinos, which requires a minimum of $n_\text{s}=3$. For our model with two massive neutrinos, Eq.\ (\ref{eq. neutrino masses}) can be inverted leading to the Casas-Ibarra parameterization \cite{Casas:2001sr}
\begin{align}
  \lambda_6^{im}=U_\text{PMNS}^{ij}\sqrt{m_{\nu_j}}R^{jl}\sqrt{A_l}^{-1}O_\eta^{lm}.\label{eq. CasasIbarra}
\end{align}
$m_{\nu_j}$ are the eigenvalues of the neutrino mass matrix $M_\nu$, and $U_{\text{PMNS}}$ the PMNS matrix \cite{Zyla:2020zbs}. $R$ is a $3\times2$ matrix fulfilling the condition
\begin{align}
  RR^T=\left(\begin{matrix}
    0 & 0 & 0\\0 & 1 & 0\\0 & 0 & 1\\
  \end{matrix}\right),
\end{align}
meaning that $R$ can be parameterized as
\begin{align}
  R=\left(\begin{matrix}
    0&0\\\cos(\theta)&-\sin(\theta)\\\sin(\theta)&\cos(\theta)\\
  \end{matrix}\right)
\end{align}
by an angle $\theta$, which is allowed to take any value between 0 and $2\pi$.

\section{Spin independent and spin dependent cross sections}
\label{sec:3}

After discussing the particles in the model, as well as the generation of neutrino masses, it is now time to turn to the interactions with SM particles through scattering. In order for DM to be captured in the Sun, it needs to lose energy by scattering off nuclei inside the Sun. The spin dependent (SD) and spin independent (SI) cross sections determine the scattering rate of WIMPs on nucleons. Experiments such as \textsc{XENON1T} and \textsc{PICO-60} search for signals from the nuclear recoil of a DM-nucleon scattering, and they have put stringent bounds on the cross sections of these processes \cite{Zyla:2020zbs,Klasen:2015uma,Aprile:2018dbl,Aprile:2019dbj,Aprile:2019xxb,Amole:2019fdf}. Both the SD and SI cross sections determine the capture rate of DM in the Sun. Therefore, they have a large impact on the neutrino flux from DM annihilations. The model we consider here contains a scalar as well as a fermionic DM candidate, both of which have different scattering phenomenology.

\begin{figure}
  \centering
  \includegraphics[scale=1]{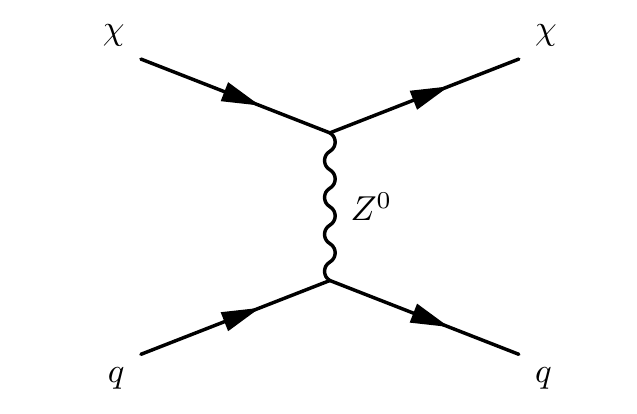}
  \includegraphics[scale=1]{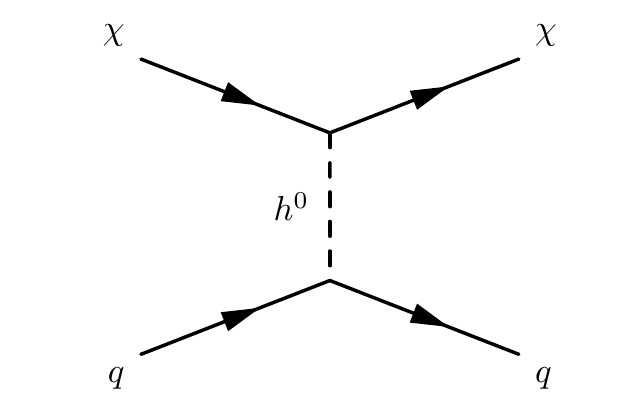}
  \caption{Feynman diagrams for DM scattering on quarks (nucleons) through $Z^0$ (left) and Higgs boson exchange (right).}
  \label{fig. scattering diagrams}
\end{figure}

In our case, scalar DM can only scatter via the SI process through the Higgs boson. The SD scattering contribution with a $Z^0$-boson as mediator is not present here, resulting in a lower cross section. As we will show later, this has a clear effect on the \textsc{IceCube} event rate. For fermionic DM there is one diagram contributing to the SI and SD cross sections each \cite{JUNGMAN1996195}. The diagram for spin (in)dependent scattering is a $t$-channel diagram with a \mbox{(Higgs-)} $Z^0$-boson as mediator as illustrated in Fig.\ \ref{fig. scattering diagrams}. Scattering processes through $s$-channel diagrams are not possible, since none of the new particles couples to quarks.
As it turns out, the SI and SD cross sections are correlated with each other, as both are dependent on the mixing between the fermion singlet and doublet, governed by $\lambda_4$ and $\lambda_5$. The explicit Feynman rules for the DM-mediator vertices as well as the impact of singlet-doublet mixing on the cross sections are discussed in App.\ \ref{app. mixing and vertices}.

\section{Detecting neutrinos from DM in the Sun with \textsc{IceCube}}
\label{sec:4}

In order to determine the event rate in \textsc{IceCube}, we start from the differential neutrino flux at Earth, which is given by 
\begin{equation}
\frac{d\phi_{\nu}}{dE_{\nu}}=\frac{1}{4\pi d_{\odot}^{2}}\Gamma_{\chi\chi}\sum_{f}Br_{f\bar{f}}\frac{dN_{f}}{dE_{\nu}}\label{eq: flux sun earth with branching}
\end{equation}
with the distance Earth-Sun $d_{\odot}$, the annihilation rate $\Gamma_{\chi\chi}$, the branching fractions $Br_{f\bar{f}}$ of final annihilation particle states $f$, and the corresponding differential neutrino energy spectra $\frac{dN_{f}}{dE_{\nu}}$. This neutrino flux is related to the differential expected signal event number by the relation
\begin{equation}
\frac{dN_{s}}{dE}=t_{e}\left(\frac{d\phi_{\nu_{\mu}}}{dE}A_{\nu_{\mu}}(E)+\frac{d\phi_{\bar{\nu}_{\mu}}}{dE}A_{\bar{\nu}_{\mu}}(E)\right)\,,\label{eq:Nsig}
\end{equation}
where $t_{e}$ is the exposure time and $A_{\nu_{\mu}(\bar{\nu}_{\mu})}$ is the muon
neutrino (muon antineutrino) effective area of the detector.  

\begin{figure}
  \centering
  \includegraphics[scale=0.5]{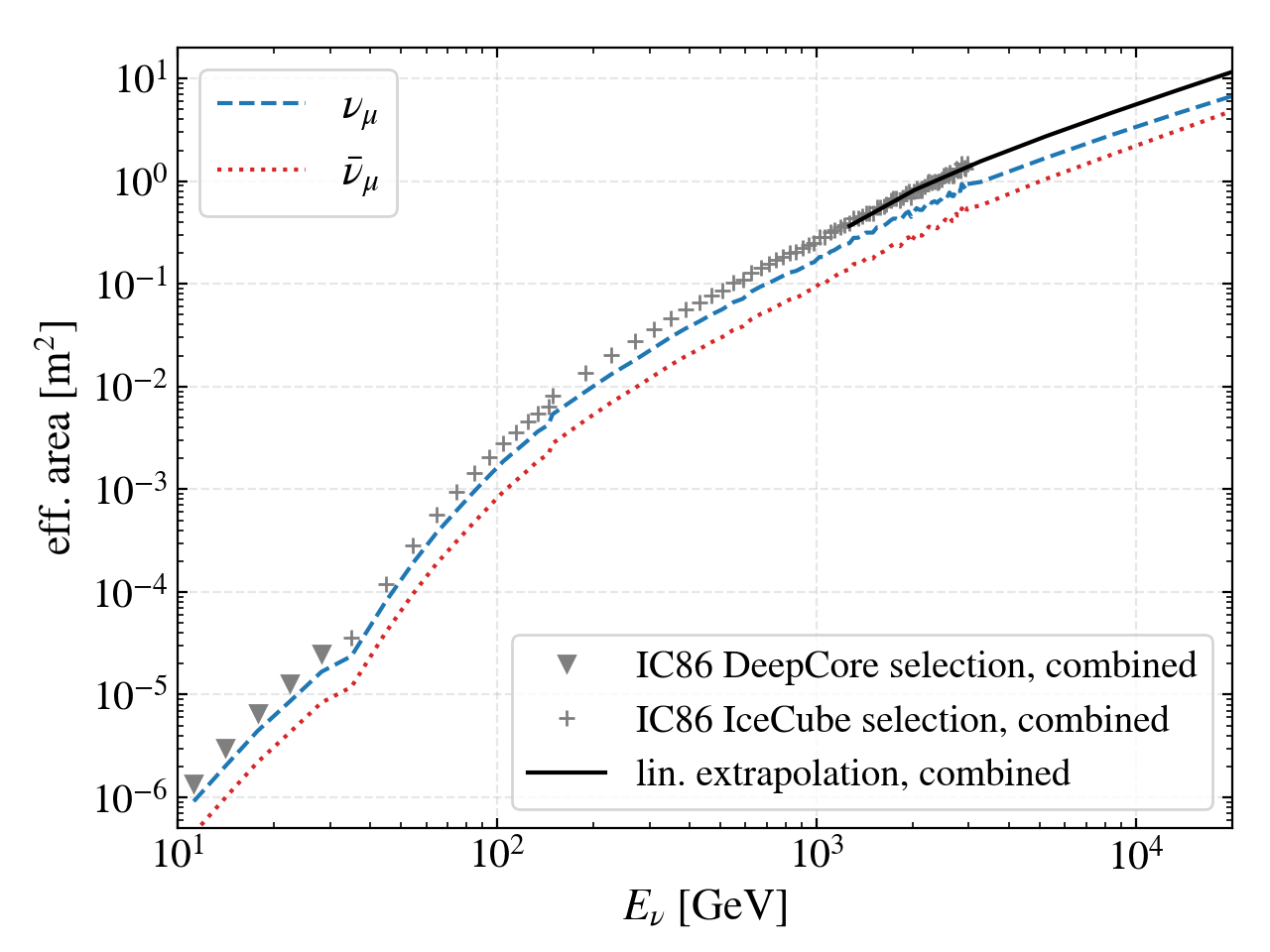}
  \caption{The $\nu_\mu$ and $\bar{\nu}_\mu$ effective areas for the \textsc{DeepCore} and the \textsc{IceCube} detector configuration \textsc{IC86} as a function of the neutrino energy. The data points are taken from Ref.\ \cite{Aartsen:2016zhm} (triangles and crosses) and linearly extrapolated (black line). Individual effective areas are calculated using data from Ref.\ \cite{Buga2003} (blue dashed and red dotted lines). Both the \textsc{IceCube} and \textsc{DeepCore} selections of the effective area are used in our work.}
  \label{fig. eff Area}
\end{figure}

To evaluate the expected event rate numerically, we implement a new routine in \textsc{micrOMEGAs} 5.0.8 \cite{Belanger:2018ccd} that computes the expected number of signals in the \textsc{IceCube} detector configuration \textsc{IC86}. A routine based on the now obsolete configuration with 22 strings (\textsc{IC22}) was already implemented, so that we could use it as a basis for our new routine. We modify it by convoluting the neutrino spectra with the effective area for \textsc{IC86} instead of \textsc{IC22}. Eight DeepCore strings are part of the \textsc{IC86} configuration. Including their effective area lowers the energy threshold to 10 GeV. We extrapolate the data points linearly to fit our energy range, as shown in Fig.\ \ref{fig. eff Area}. The corresponding data points have been taken from Ref.\ \cite{Aartsen:2016zhm}. In the region where both selections overlap, we use the \textsc{IceCube} effective area, as it is larger than the one of the \textsc{DeepCore} selection.

For deep-inelastic scattering of neutrinos with nucleons, the energy-dependent cross sections for $\nu_{\mu}$ and $\bar{\nu}_{\mu}$ are different. The combined effective area was split up into individual effective areas by taking these differences into account and with data from Ref.\ \cite{Buga2003} and the relation
\begin{equation}
A_{\nu_{\mu}\left(\bar{\nu}_{\mu}\right)}=\frac{A_{\text{combined}}}{1+\frac{\sigma_{\bar{\nu}_{\mu}\left(\nu_{\mu}\right)}}{\sigma_{\nu_{\mu}\left(\bar{\nu}_{\mu}\right)}}}\,.\label{eq: Aeff}
\end{equation}

In Tab.\ \ref{tab. Benchmark} we show the parameters for a typical benchmark point.
\begin{table}
  \centering
  \caption{Parameters used for our benchmark point (masses in GeV). The matrix elements not mentioned (e.g. $\lambda_1^{12}$) are set to zero.}
  \label{tab. Benchmark}
  \begin{tabular}{|cc|cc|}
  \hline
    $M_\Psi$ & $M_{\psi\psi'}$ & $(M_\phi^2)_{11}$ & $(M_\phi^2)_{22}$ \\
    \hline
    $362$ & $614$ & $2.4\cdot 10^{6}$ & $4.3\cdot 10^{7}$ \\
    \hline
  \end{tabular}
  \vspace{5mm}
  \begin{tabular}{|cc|cc|cccccc|}
    \hline
    $\lambda_4$ & $\lambda_5$ & $\lambda_1^{11}$ & $\lambda_1^{22}$ & $\lambda_6^{11}$ & $\lambda_6^{12}$ & $\lambda_6^{21}$ & $\lambda_6^{22}$ & $\lambda_6^{31}$ & $\lambda_6^{32}$\\
    \hline
    $-0.17$ & $-0.65$ & $0.011$ & $0.012$ & $-1.74\cdot 10^{-5}$ & $-5.28\cdot 10^{-5}$ & $0.90\cdot 10^{-5}$ & $-18.67\cdot 10^{-5}$ & $5.36\cdot 10^{-5}$ & $11.91\cdot 10^{-5}$\\
    \hline
  \end{tabular}
\end{table}
\begin{figure}
  \centering
  \includegraphics[scale=0.5]{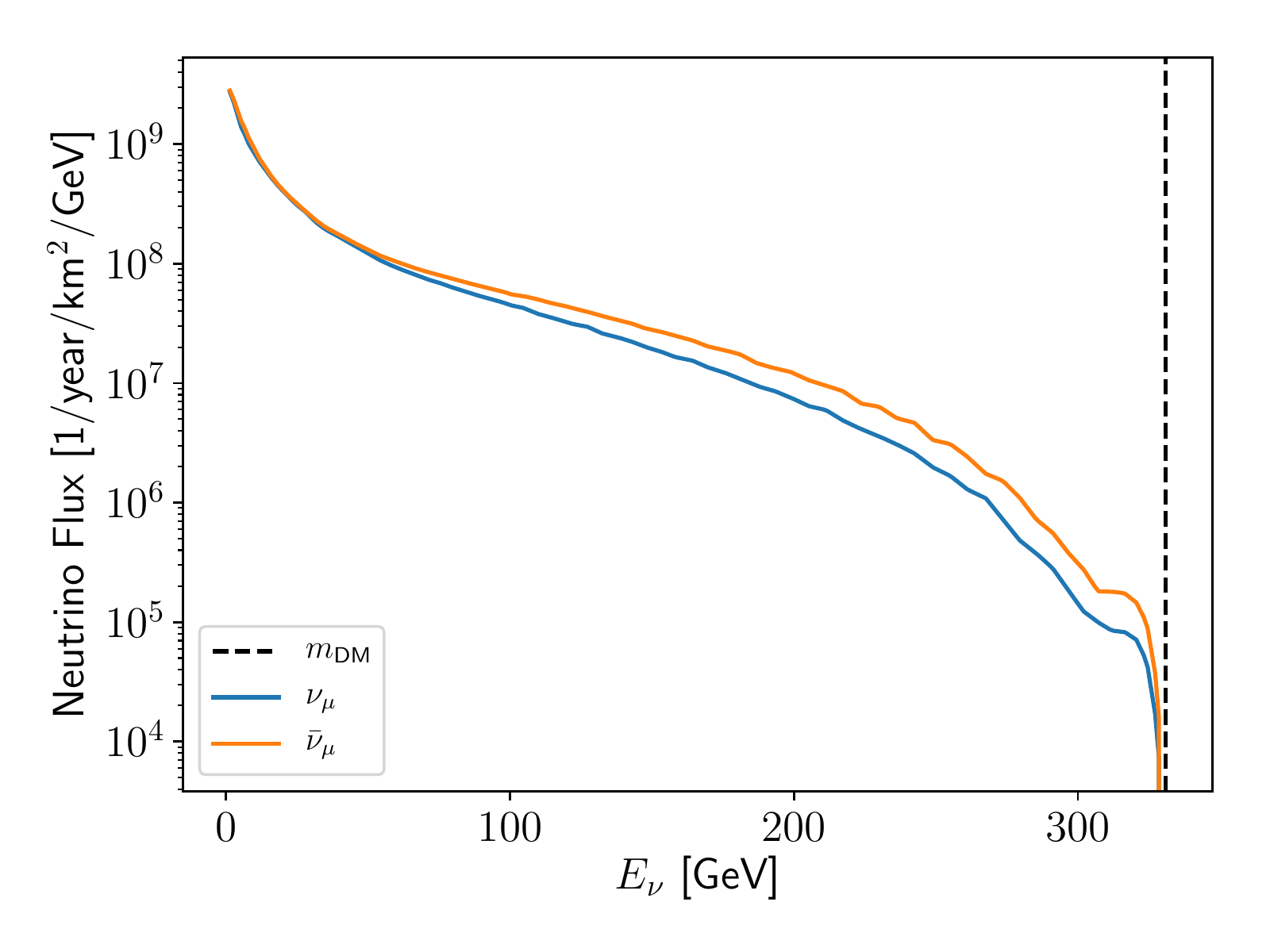}
  \includegraphics[scale=0.5]{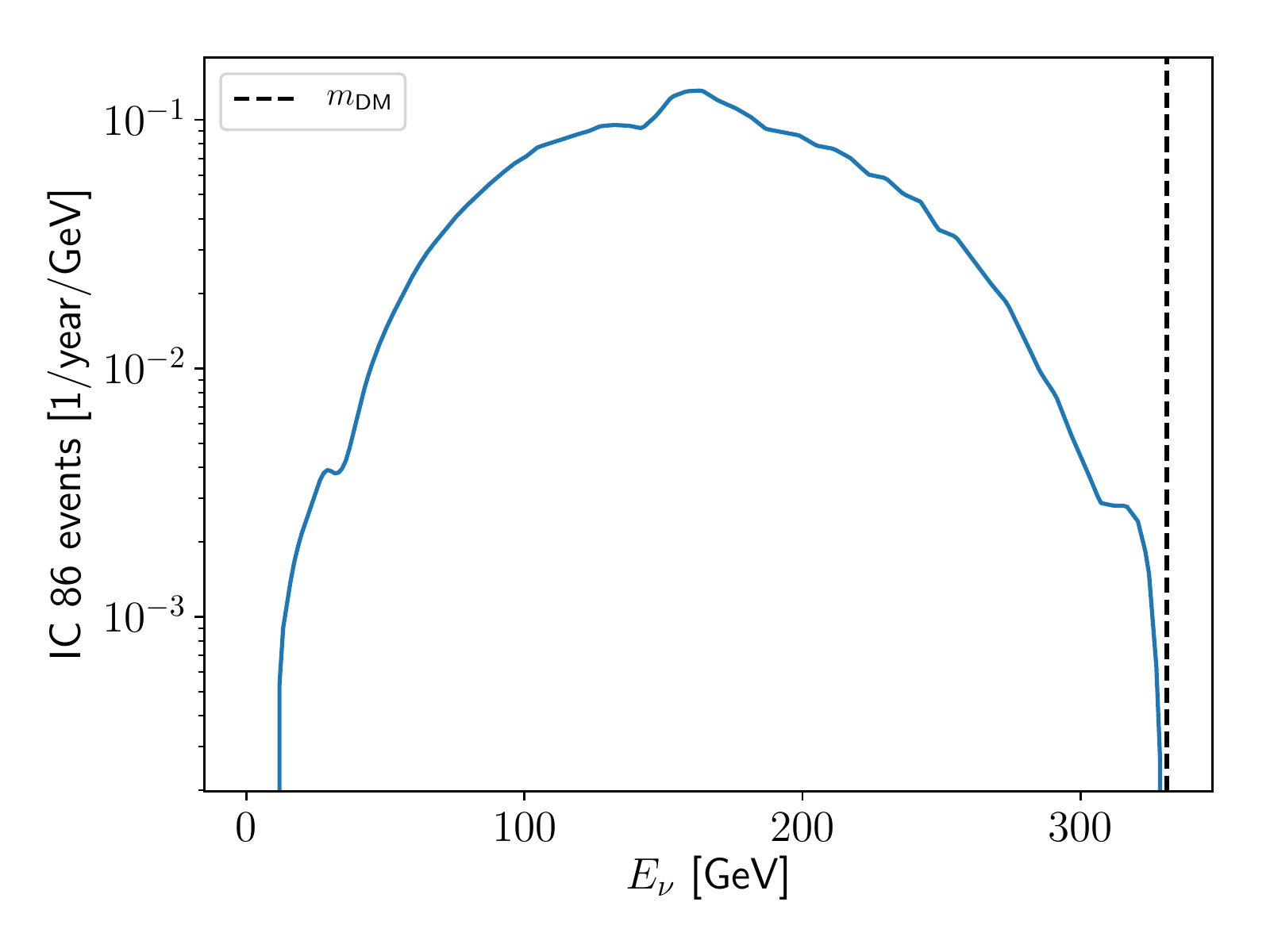}
  \caption{(Anti-)neutrino flux from the Sun (left) and expected event rate at \textsc{IceCube} (right) for the benchmark point given in Tab.\ \ref{tab. Benchmark}. The dashed lines indicate the value of the DM mass.}
  \label{fig. BS NuFlux Events}
\end{figure}
This benchmark point gives an expected event rate at \textsc{IceCube} of $16$ events per year. Fig.\ \ref{fig. BS NuFlux Events} shows the differential neutrino and antineutrino fluxes, which differ due to absorption (also oscillation and regeneration) effects in the Sun \cite{Cirelli:2005gh}, and the expected event rate at \textsc{IceCube} at different energies. The neutrino flux is highest for energies below 50 GeV, but even when taking into account the DeepCore effective area from Fig.\ \ref{fig. eff Area}, the amount of events from these neutrinos is suppressed by the small effective area. There is no neutrino flux for $E_\nu>m_\text{DM}$, since the neutrinos cannot have more kinetic energy than the mass of the annihilating particle. The velocities of the annihilating DM particles are negligible. For a typical point such as this benchmark point, there are almost no monochromatic neutrinos with $E_\nu=m_\text{DM}$. 
Thus we do not expect to measure a neutrino line at \textsc{IceCube}. It turns out that direct annihilation of fermion DM into a SM neutrino line is also suppressed in general by the Yukawa couplings $\lambda_6$, restricted to be small by the neutrino masses and LFV (see below) \cite{Fiaschi:2018rky}. However, a continuous signal spectrum as shown in Fig.\ \ref{fig. BS NuFlux Events} is indeed expected from the Sun. For the benchmark point chosen above, the SI cross section is $\sigma_p\text{(SI)}=1.75\cdot10^{-8}$ pb and already excluded by the \textsc{XENON1T} limits, but as we will see in the following, this statement does not generalize.

\section{Numerical results}
\label{sec:5}

We explore the parameter space of our model by means of a numerical scan. We use \textsc{SPheno} 4.0.3 \cite{Porod:2003um,Porod:2011nf} to calculate the masses of the new particles and the branching ratios (BRs) for Lepton Flavor Violation (LFV). The \textsc{SPheno} module is generated using \textsc{SARAH} 4.14.0 \cite{Staub:2013tta}. The relic density, spin (in)dependent cross sections and the neutrino flux from the Sun are subsequently calculated using \textsc{micrOMEGAs} 5.0.8 \cite{Belanger:2018ccd}. 

We now scan over the parameter space of the model. The mass parameters $M_{\Psi,\psi\psi',\phi}$ are varied between $100\text{ GeV}$ and $10\text{ TeV}$. As was mentioned before, $M_\phi^2$ is chosen to be diagonal. The couplings $|\lambda_{1,4,5}|$ are varied between $1\cdot10^{-3}$ and $10$, where $\lambda_1$ is chosen to be diagonal. The signs of these couplings are chosen randomly. $\lambda_6$ is calculated using the Casas-Ibarra parameterization Eq.\ (\ref{eq. CasasIbarra}), which requires the other model parameters as input. The neutrino mass differences and the PMNS angles are varied in the $3\sigma$ ranges \cite{Esteban:2018azc}, where we assume Normal Ordering. The angle $\theta$ from the Casas-Ibarra parameterization is varied from zero to $2\pi$. We require the relic density to be $\Omega h^2=0.12$ \cite{Aghanim:2018eyx}, allowing it to vary by $\pm 0.02$. In addition we require the Higgs mass to be $(125\pm2.5)$ GeV. Lepton Flavor Violation (LFV) constrains the parameter space further. We impose the current limits on BR$(\mu\to e\gamma)<4.2\cdot10^{-13}$ \cite{TheMEG:2016wtm} and BR$(\mu\to 3e)<1.0\cdot10^{-12}$ \cite{BELLGARDT19881}, as they impose usually the most stringent constraints. As a cross check, we have reproduced the results shown in Figs.\ 7 and 10 of Ref.\ \cite{Fiaschi:2018rky}.

\subsection{Spin independent scattering}

\begin{figure}
  \centering
  \includegraphics[scale=0.75]{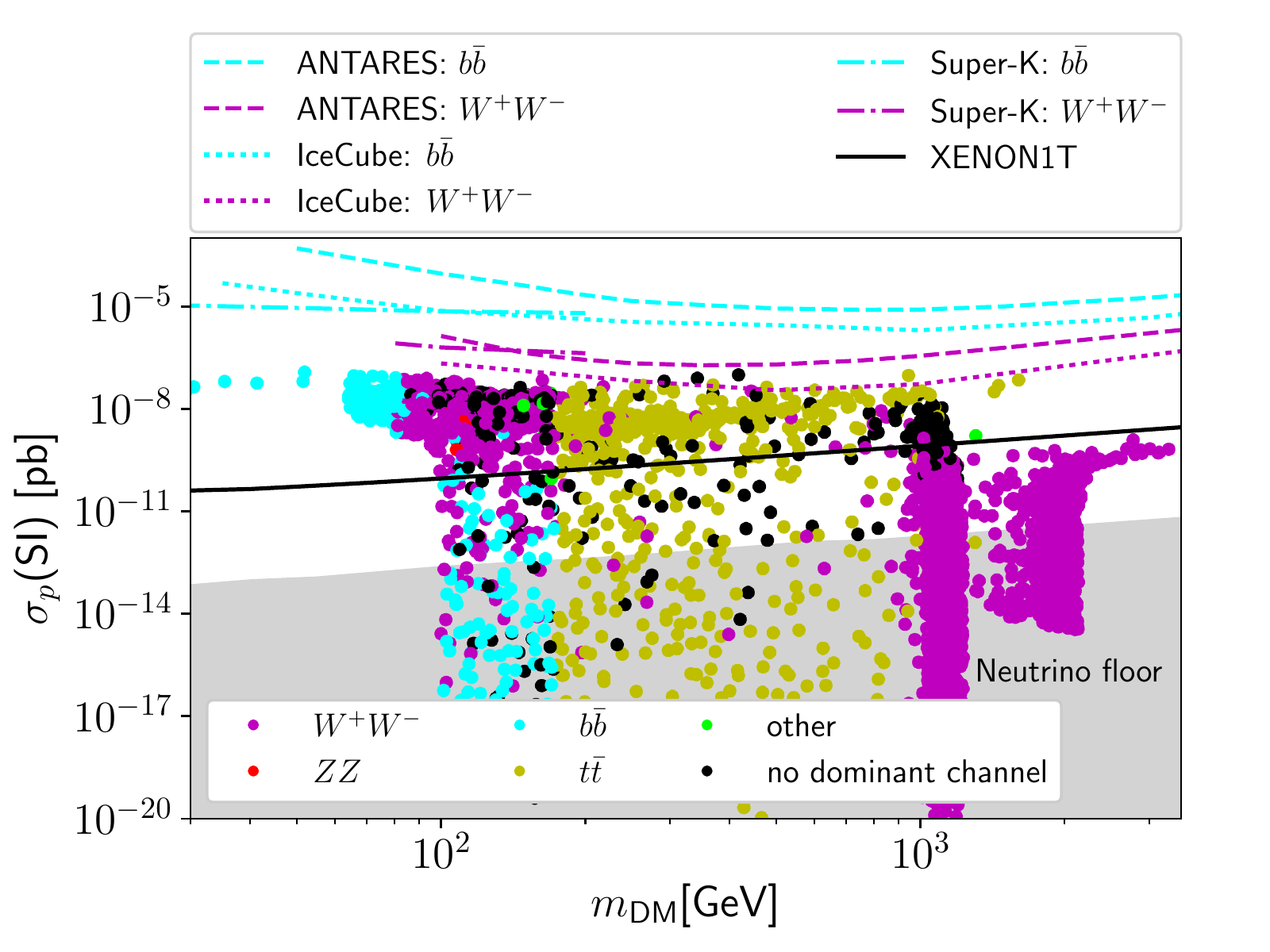}
  \caption{The spin independent (SI) cross section in pb with \textsc{ANTARES} \cite{Adrian-Martinez:2016gti}, \textsc{IceCube} \cite{Aartsen:2016zhm}, \textsc{Super-Kamiokande} \cite{Choi:2015ara} and \textsc{XENON1T} \cite{Aprile:2018dbl} exclusion limits as a function of the DM mass for both fermion (below about 1 TeV) and scalar DM (above). All points and lines are color coded according to the main annihilation channel, provided there is one with a branching ratio of over 50\%. Also shown is the neutrino floor \cite{Billard:2013qya}.}
  \label{fig05}
\end{figure}

Our spin independent cross section results for both fermion (below about 1 TeV) and scalar (above about 1 TeV) DM are shown in Fig.\ \ref{fig05}. Taking into account a slightly different spread of the points due to differing scan ranges and the already imposed LFV limits, the results agree with those in Figs.\ 7 and 10 of Ref.\ \cite{Fiaschi:2018rky}. The sampled points are color coded according to the dominant annihilation branching ratio. A point is marked as having no dominant channel when no single branching ratio reaches 50\%.

The scalar DM candidates, all located around 2 TeV, mainly decay into a pair of $W$-bosons. For fermionic DM the situation is mixed. At masses below the $W$ boson mass, the main channel is through $b \bar{b}$ production, after which the dominant channel becomes $W^+ W^-$. For masses above the top mass, there is mainly $t \bar{t}$ production, except for the points around 1 TeV, where $W$-boson production is the only dominant channel. This can be explained by the singlet-doublet nature of the fermionic DM, where the parameter points located around 1 TeV are those that have a large doublet contribution and therefore couple more strongly to the electroweak gauge bosons. In contrast, the points for which the DM candidate is mainly a singlet with only a small doublet admixture couple less to the gauge bosons and relatively more to the top quark via the SM Higgs boson. The charged fermions are always heavier than 102 GeV, so that the limits by the LEP experiments \cite{Abbiendi:2003yd} do not restrict the parameter space. We see that the previous limits by the \textsc{ANTARES} \cite{Adrian-Martinez:2016gti}, \textsc{IceCube} \cite{Aartsen:2016zhm} and \textsc{Super-Kamiokande} \cite{Choi:2015ara} collaborations from DM annihilations in the Sun are several orders of magnitude weaker than the \textsc{XENON1T} \cite{Aprile:2018dbl} direct detection bound, with the $b \bar{b}$ limits being less stringent compared to the $W^+ W^-$ ones. We also show the atmospheric and diffuse supernova background (DSNB) neutrino ``floor'' \cite{Billard:2013qya}, which may render DM direct detection difficult.

\subsection{Spin dependent scattering}

\begin{figure}
  \centering
  \includegraphics[scale=0.75]{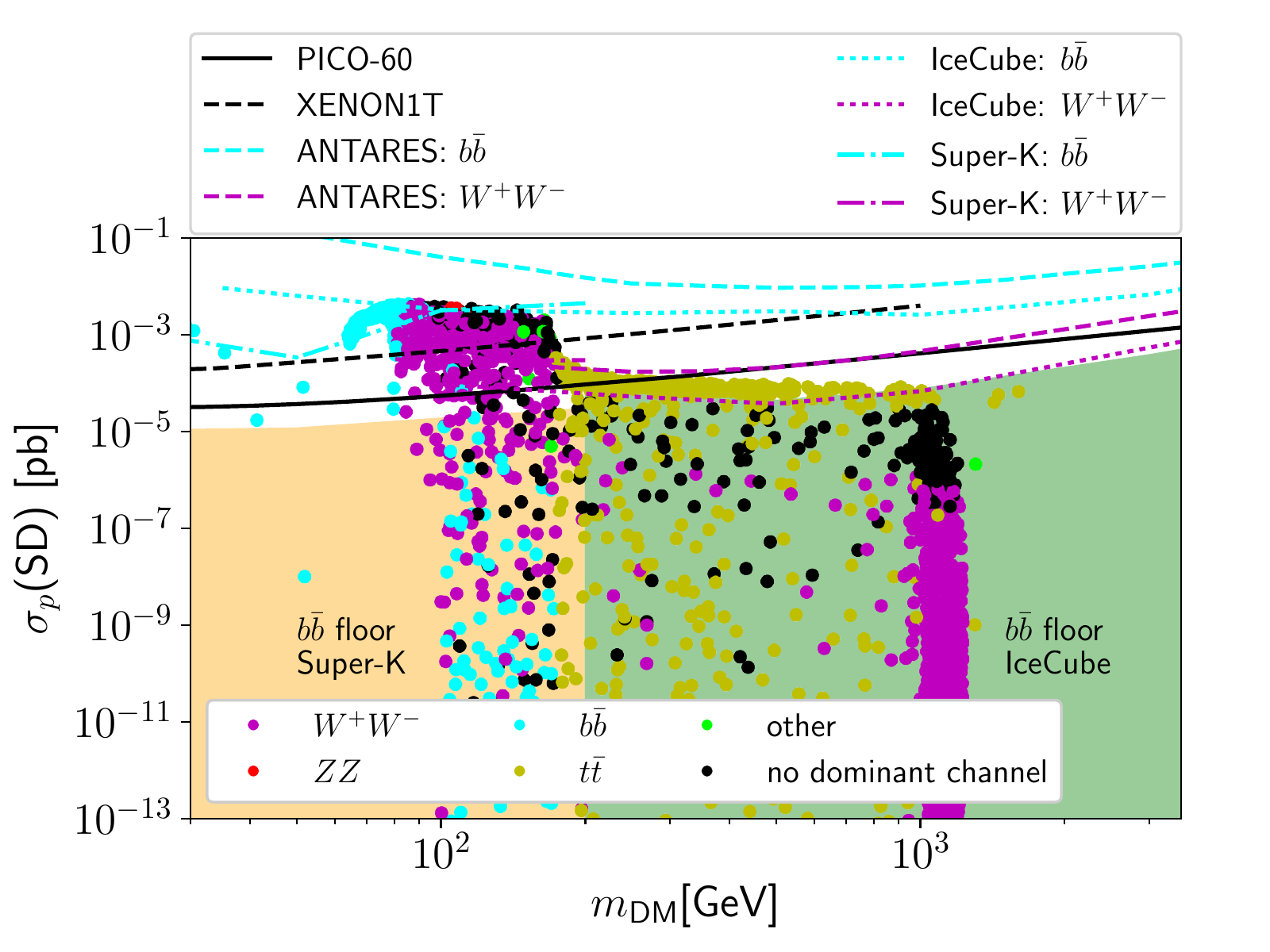}
  \caption{The spin dependent (SD) cross section in pb with \textsc{PICO-60} \cite{Amole:2019fdf}, \text{XENON1T} \cite{Aprile:2019dbj}, \text{ANTARES} \cite{Adrian-Martinez:2016gti}, \textsc{IceCube} \cite{Aartsen:2016zhm} and \textsc{Super-Kamiokande} \cite{Choi:2015ara} exclusion limits as a function of the DM mass for singlet-doublet fermion DM. All points and lines are color coded according to the main annihilation channel, provided there is one with a branching ratio of over 50\%. Also shown is the neutrino floor for $b\bar{b}$ final states \cite{Ng:2017aur}.}
  \label{fig06}
\end{figure}

In Fig.\ \ref{fig06} we show the spin dependent cross section for singlet-doublet fermion DM in our model and compare it to direct and indirect detection limits. Different annihilation channels are again color coded as before. For certain channels, \text{ANTARES} \cite{Adrian-Martinez:2016gti} and \textsc{IceCube} \cite{Aartsen:2016zhm} impose stronger constraints than \text{XENON1T} \cite{Aprile:2019dbj}, while the full data set of \textsc{PICO-60} \cite{Amole:2019fdf} has similar sensitivity as the indirect detection experiments. In a combined analysis, \textsc{IceCube} and \textsc{PICO-60} have removed the Standard Halo Model assumption and published velocity independent limits following the suggestion in Ref.\ \cite{Ferrer:2015bta}, which are however significantly weaker \cite{Amole:2019coq}. Also shown is the neutrino floor for $b\bar{b}$ final states due to high-energy neutrinos from cosmic-ray interactions with the solar atmosphere, which may render indirect detection difficult \cite{Ng:2017aur} and leaves little room for the $b\bar{b}$ channel at low mass beyond the \textsc{PICO-60} limits.

\subsection{Limits from the Galactic Center}

\begin{figure}
  \centering
  \includegraphics[scale=0.75]{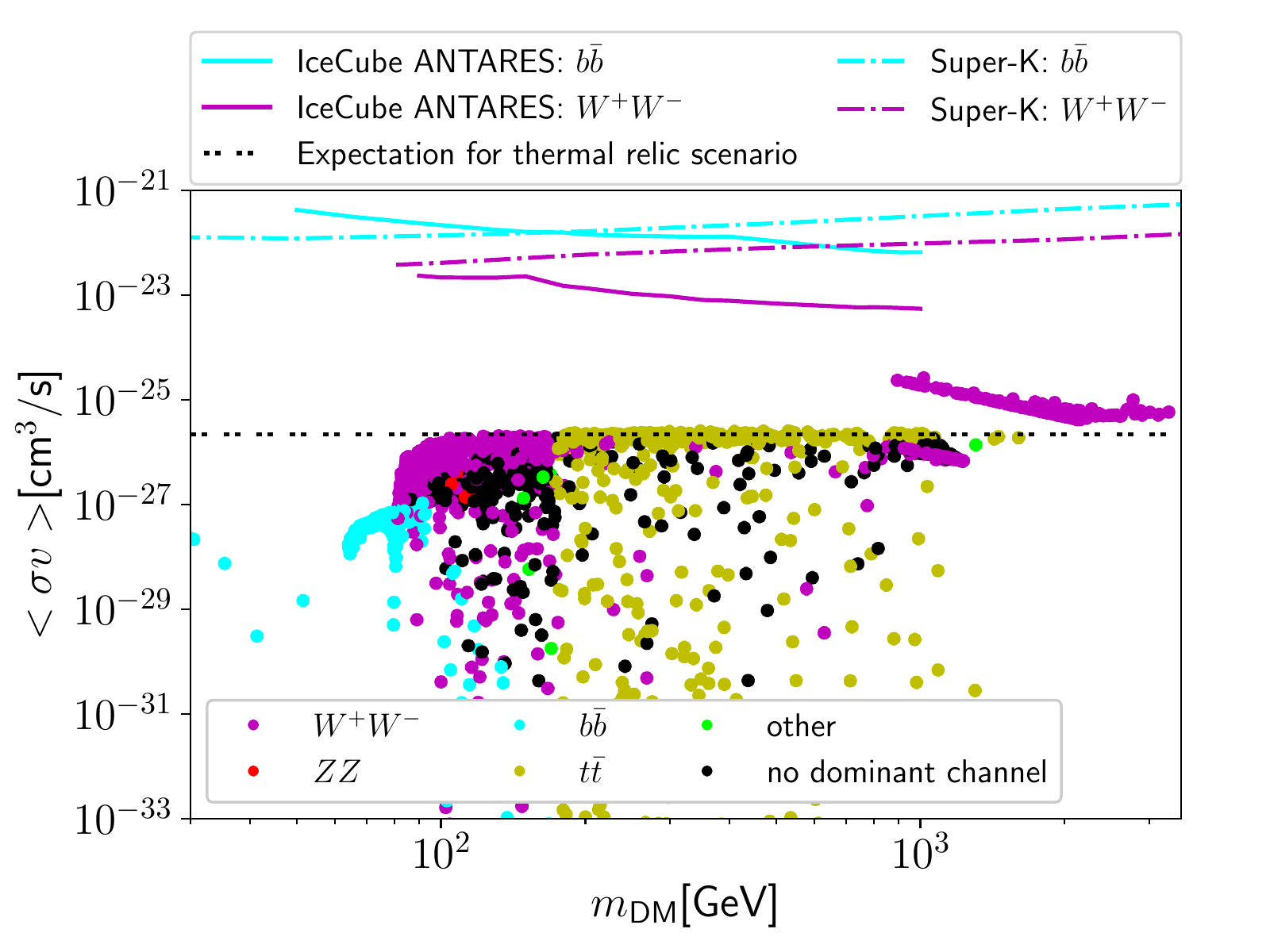}
  \caption{Thermally averaged cross section $\langle \sigma v \rangle$ with combined \textsc{IceCube} \textsc{ANTARES} \cite{Aartsen:2020tdl} and \textsc{Super-Kamiokande} \cite{Abe:2020sbr} exclusion limits as a function of the DM mass for both fermion (below about 1 TeV) and scalar DM (above). All points are colored according to the main annihilation channel, provided there is one with a branching ratio of over 50\%. Also shown is the expected cross section for a thermal relic \cite{Steigman:2012nb}.}
  \label{fig07}
\end{figure}

Fig.\ \ref{fig07} shows the thermally averaged cross section $\langle \sigma v \rangle$ for the same points with fermionic and scalar DM as in Fig.\ \ref{fig05}. Here, we assume the NFW DM halo profile \cite{Navarro:1995iw}. The expectation for a thermal relic is indicated by a dashed line \cite{Steigman:2012nb}. For fermionic DM, it represents an upper limit, while for scalar DM it is rather a lower limit. All points are several orders of magnitude below the bounds established by \textsc{IceCube} \cite{Aartsen:2017ulx}, \textsc{ANTARES} \cite{Albert:2016emp} and their combination \cite{Aartsen:2020tdl} as well as by \textsc{Super-Kamiokande} \cite{Abe:2020sbr}, meaning that these measurements do not constrain the model. For heavy scalar DM, the \textsc{IceCube} \textsc{ANTARES} sensitivity must be improved less than for lighter fermion DM, i.e.\ by only a few orders of magnitude.

\subsection{Expected IceCube event rates}

\begin{figure}
  \centering
  \includegraphics[scale=0.75]{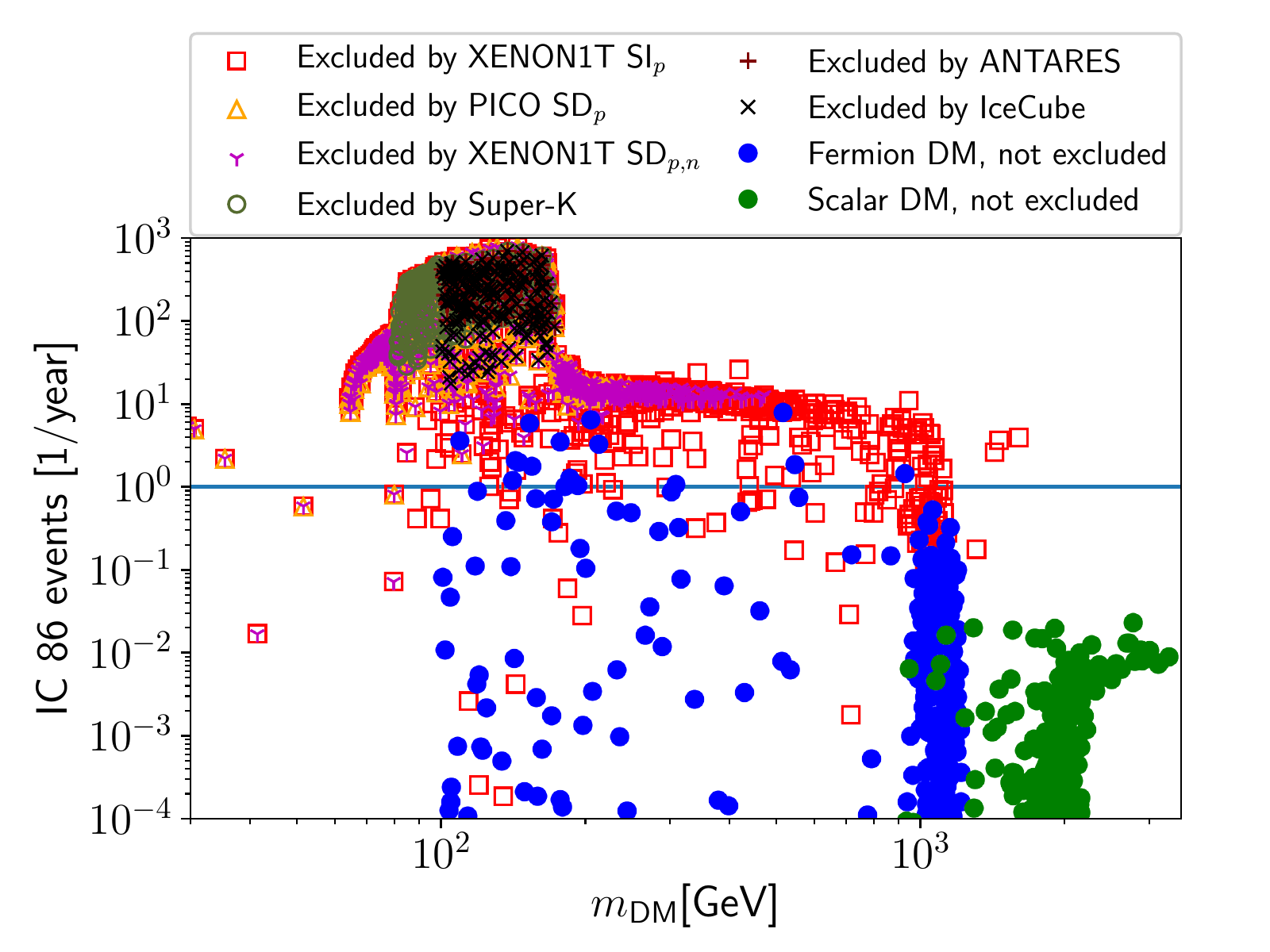}
  \caption{The expected number of events per year in the current \textsc{IceCube} configuration with 86 strings (\text{IC86}) as a function of the DM mass for singlet-doublet fermion (blue) and triplet scalar (green) DM. Allowed points are shown together with points excluded by direct and indirect detection (other colors and symbols). The blue line marks one event per year for orientation.}
  \label{fig08}
\end{figure}

In Fig.\ \ref{fig08} we show the expected event rate at \textsc{IceCube} for all model points of our numerical scan. For the triplet scalar DM case (green points), the spin dependent cross section is always zero. Thus in our model the accumulation of scalar DM in the Sun is only determined by the spin independent cross section, which lies below the current \textsc{XENON1T} bound. As can be seen, this leads to less than one event per year in the current \textsc{IceCube} configuration with 86 strings (\textsc{IC86}), whose sensitivity would therefore have to be improved by a few orders of magnitude.

For singlet-doublet fermion DM (blue and other points), the expected event rates reach values of up to 1000 events per year. However, we have to impose all previous direct and indirect detection constraints, marked by different symbols and colors in Fig.\ \ref{fig08}. Not shown previously, but also imposed are the \textsc{XENON1T} limits on spin dependent scattering off neutrons, which occurs rarely in the Sun. For indirect detection, we have always used the limits for the main annihilation channel. As expected, viable models (blue) lie in particular below the rates excluded previously by \text{IceCube} (black x symbols). They can reach rates of up to ten events per year at \textsc{IC86}, making indirect detection competitive with respect to the direct detection limits imposed in particular by \text{PICO-60} (orange triangles). A considerable fraction of the parameter space with high rates is excluded by the limits on the spin independent cross section set by \textsc{XENON1T} (red squares) due to the correlation of the spin dependent and spin independent cross sections through $\lambda_4$ and $\lambda_5$ (see App.\ \ref{app. mixing and vertices}). Note, however, that the correlation of spin dependent and spin independent cross sections is absent in fine-tuned scenarios where the relation between $\lambda_4$ and $\lambda_5$ is fixed as described in the Appendix. 


\section{Summary and outlook}
\label{sec:6}

To summarize, we have studied in this paper the prospects to probe radiative seesaw models with neutrino signals from DM annihilation and detectors such as \textsc{ANTARES}, \textsc{Super-Kamiokande} and in particular \textsc{IceCube}, focusing on the model T1-3-B with $\alpha=0$ with either scalar triplet or singlet-doublet fermion DM. Both DM candidates can in principle directly annihilate into neutrinos. However, the relevant Yukawa couplings involved are usually strongly constrained to be small from neutrino masses and LFV processes, which are always present in these models. A sharp neutrino line at an energy corresponding to the DM mass is therefore not expected.

Continuous neutrino spectra are, however, produced from DM annihilation into decaying SM particles such as $W$ and $Z$ bosons, $b$ and $t$ quarks as well as (at least in principle) $\mu$ and $\tau$ leptons. When boosted through DM accumulation in celestial bodies such as the Earth, the Sun or the Galactic Center, the rates are observable in neutrino telescopes. Focusing on the most promising case of the Sun, we performed a detailed analysis of the expected event rates in \textsc{IceCube}. In the case of scalar triplet DM, there exists no spin dependent scattering process, and the spin independent scattering cross section is too small to obtain enough accumulation inside the Sun. Because of this, the event rate of neutrino signals in \textsc{IceCube} would lie below one event per year, making scalar triplet DM currently undetectable with neutrino telescopes.

For singlet-doublet fermionic DM, the situation is different. In this scenario, the DM candidate can scatter via both the spin independent as well as the spin dependent process, leading to rates of up to 1000 events per year. Through our approximation of the fermion mixing matrix for small Yukawa couplings $\lambda_{4,5}$, we showed then that the DM-mediator vertices in the spin (in)dependent processes, i.e.\ with $Z^0$ (Higgs) bosons, both depend on these Yukawa couplings, so that the two scattering processes become correlated. This was confirmed in a numerical scan and resulted in a considerable fraction of the parameter space with large event rates being excluded by the stringent \textsc{XENON1T} limits on the spin independent cross section. Previously obtained results by \textsc{ANTARES}, \textsc{IceCube} and \textsc{Super-Kamiokande} from the Sun and the Galactic Center were instead found to be much weaker. Constraints on the spin dependent cross section from \textsc{PICO-60} and previous analyses by \textsc{IceCube} and \textsc{ANTARES} limited the viable models further to event rates of up to ten per year, leaving indirect detection with the \textsc{IceCube} neutrino telescope still competitive with respect to direct detection experiments.

Our results generalize to models with either real scalar triplet DM or fermion DM with singlet doublet mixing only, where scattering in the Sun is governed by similar relationships.


\begin{acknowledgments}
We thank Carsten Rott for useful comments on the manuscript. This work has been supported by the DFG through the Research Training Network 2149 ``Strong and weak interactions - from hadrons to dark matter'' and by BMBF through Verbundforschung grant 05A20PM2.
\end{acknowledgments}

\appendix

\section{Analytic results for singlet-doublet fermion mixing and vertices}
\label{app. mixing and vertices}

The spin (in)dependent cross sections mainly depend on the three point vertex between two DM particles and the (Higgs-) $Z^0$-boson. We focus on the fermion DM case since the spin dependent cross section for a scalar triplet is always zero and its coupling to the Higgs boson is given by $\lambda_1$. The vertices for the mass eigenstates of the fermions can be calculated by \textsc{SARAH} \cite{Staub:2013tta}:
\begin{align}
\raisebox{-2.5em}{\includegraphics[scale=1]{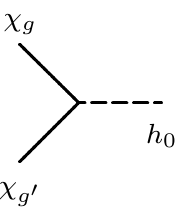}}=&-\frac{1}{\sqrt{2}}i \left\{(U_\chi)^*_{g1} \left[\lambda_4
   (U_\chi)^*_{g'3}+\lambda_5
   (U_\chi)^*_{g'2}\right]+\lambda_4
   (U_\chi)^*_{g3}
   (U_\chi)^*_{g'1}+\lambda_5
   (U_\chi)^*_{g2}
  (U_\chi)^*_{g'1}\right\}P_\text{L}\nonumber\\
    &-\frac{1}{\sqrt{2}}i \left\{(U_\chi)_{g1} \left[\lambda_4
   (U_\chi)_{g'3}+\lambda_5
   (U_\chi)_{g'2}\right]+\lambda_4 (U_\chi)_{g3}
   (U_\chi)_{g'1}+\lambda_5 (U_\chi)_{g2}
(U_\chi)_{g'1}\right\}P_\text{R},\\
\raisebox{-2.5em}{\includegraphics[scale=1]{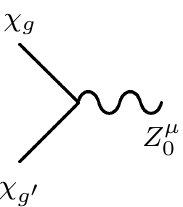}}
=&-\frac{1}{2} i \left[g_1\sin(\theta_\text{W})+g_2\cos(\theta_\text{W})\right]\left[(U_\chi)_{g2}(U_\chi)^*_{g'2}-(U_\chi)_{g3}(U_\chi)^*_{g'3}\right]\gamma^\mu P_\text{L}\nonumber\\
   &+\frac{1}{2} i \left[g_1\sin(\theta_\text{W})+g_2\cos(\theta_\text{W})\right]\left[(U_\chi)_{g'2}(U_\chi)^*_{g2}-(U_\chi)_{g'3}(U_\chi)^*_{g3}\right] \gamma^\mu P_\text{R}.
\end{align}
Since the three mass eigenstates are generally not mass degenerate, only the elastic case where $g=g'$, with $\chi_g$ being DM, contributes to the spin (in)dependent cross section. Both vertices depend on the mixing matrix $U_\chi$ between the fermion singlet and doublet. Since the singlet-doublet mixing is induced by $\lambda_{4,5}$ as can be seen from the fermionic mass matrix in Eq.\ (\ref{eq. fermion masses}), the mixing matrix $U_\chi$ depends on $\lambda_{4,5}$.

Diagonalizing the fermionic mass matrix proves to be difficult and gives quite unwieldy results. We can however, expand the problem for small $\lambda_{4}$,
\begin{align}
  M_f=M_0+\lambda_4M_\lambda=\left(\begin{matrix}
    M_\Psi & 0&0 \\
    0 & 0& M_{\psi\psi'} \\
    0 & M_{\psi\psi'} & 0
  \end{matrix}\right)+\lambda_4\left(\begin{matrix}
    0 & \frac{\tilde{\lambda} v}{\sqrt{2}} & \frac{v}{\sqrt{2}} \\
    \frac{\tilde{\lambda} v}{\sqrt{2}} & 0& 0 \\
    \frac{ v}{\sqrt{2}} &0 & 0
  \end{matrix}\right)
\end{align}
and then (keeping $\tilde{\lambda}=\frac{\lambda_5}{\lambda_4}$ fixed) use perturbation theory to diagonalize this matrix approximately. At second order in $\lambda_4$ we obtain for the mixing matrix

\begin{align}
&U_\chi=\left(\begin{matrix}
  1&0&0\\0&-\frac{1}{\sqrt{2}}&\frac{1}{\sqrt{2}}\\0&\frac{1}{\sqrt{2}}&\frac{1}{\sqrt{2}}
\end{matrix}\right)+\\
&\left(
\begin{matrix}
 v^2\frac{\left(\lambda_5^2+\lambda_4^2  \right)\left(M_\Psi^2+M_{\psi\psi'}^2\right)+4\lambda_4 \lambda_5 M_\Psi M_{\psi\psi'}}{4 (M_\Psi-M_{\psi\psi'})^2 (M_\Psi+M_{\psi\psi'})^2}&
  -\frac{\left(\lambda_4-\lambda_5\right) v}{2 (M_\Psi+M_{\psi\psi'})}&
  -\frac{\left(\lambda_4+\lambda_5\right) v}{2 (M_\Psi-M_{\psi\psi'})}\\
 \frac{\left(\lambda_4 M_{\psi\psi'}+\lambda_5 M_\Psi\right) v}{\sqrt{2} (M_\Psi-M_{\psi\psi'}) (M_\Psi+M_{\psi\psi'})}&
  v^2\frac{\left(\lambda_4^2 -\lambda_5^2\right) M_\Psi+2\lambda_4 M_{\psi\psi'}\left(\lambda_4 -\lambda_5\right)}{8 \sqrt{2}M_{\psi\psi'} \left(M_\Psi+M_{\psi\psi'}\right)^2} &
 v^2\frac{\left(\lambda_4^2 -\lambda_5^2\right) M_\Psi-2\lambda_4 M_{\psi\psi'}\left(\lambda_4 +\lambda_5\right)}{8 \sqrt{2}M_{\psi\psi'} \left(-M_\Psi+M_{\psi\psi'}\right)^2} \\
 \frac{\left(\lambda_4 M_\Psi+\lambda_5 M_{\psi\psi'}\right) v}{\sqrt{2} (M_\Psi-M_{\psi\psi'}) (M_\Psi+M_{\psi\psi'})}&
 v^2\frac{\left(\lambda_4^2 -\lambda_5^2\right) M_\Psi+2\lambda_5 M_{\psi\psi'}\left(\lambda_4 -\lambda_5 \right)}{8 \sqrt{2}M_{\psi\psi'} \left(M_\Psi+M_{\psi\psi'}\right)^2} & 
 v^2\frac{\left(\lambda_5^2 -\lambda_4^2\right) M_\Psi-2\lambda_5 M_{\psi\psi'}\left(\lambda_5 +\lambda_4\right)}{8 \sqrt{2}M_{\psi\psi'} \left(-M_\Psi+M_{\psi\psi'}\right)^2}\\
\end{matrix}
\right)
\end{align}
and for the diagonalized mass matrix
\begin{align}
\left(
\begin{matrix}
 M_\Psi+ \frac{v^2\left(\lambda_5^2 M_\Psi+2\lambda_4
   \lambda_5 M_{\psi\psi'}+\lambda_4^2 M_\Psi\right)}{2(M_\Psi-M_{\psi\psi'}) (M_\Psi+M_{\psi\psi'})} & 0 & 0 \\
 0 &-M_{\psi\psi'} -\frac{(\lambda_4-\lambda_5)^2 v^2}{2 (M_\Psi+M_{\psi\psi'})}& 0 \\
 0 & 0 & M_{\psi\psi'}-\frac{(\lambda_4+\lambda_5)^2 v^2}{2 (M_\Psi-M_{\psi\psi'})}\\
\end{matrix}
\right).
\end{align}
One can see that in the case $M_\Psi<M_{\psi\psi'}$ (assuming $\lambda_{4,5}^2v^2<M_{\psi\psi'}^2-M_\Psi^2$) \footnote{Otherwise a more careful analysis is required. E.g. if $\lambda_4\lambda_5>0$ or $\lambda_{4,5}<\lambda_{5,4}\frac{M_\Psi}{M_{\psi\psi'}}$, the statement above is also true.}, $\chi_1$ is the lightest fermion whereas in the case $M_{\psi\psi'}<M_\Psi$, $\chi_3$ is the lightest one. Now we can use the result for $U_\chi$ to calculate how the vertices depend on $\lambda_{4,5}$ and the mass parameters $M_\Psi,M_{\psi\psi'}$. We expand the results again for small $\lambda_{4,5}$, omitting all terms that contain higher orders than $\lambda_4^2,\lambda_5^2$ or $\lambda_4\lambda_5$. The results are:
\begin{align}
  \chi_1\chi_1h^0:\quad& -\frac{iv\left( M_\Psi  \lambda_5^2+2 M_{\psi\psi'} \lambda_5 \lambda_4+M_\Psi \lambda_4^2\right)}{M_\Psi^2-M_{\psi\psi'}^2}+\mathcal{O}\left(\lambda_{4,5}^3\right),\label{eq. vertex c1c1h0}\\
  \chi_2\chi_2h^0:\quad& \frac{i v \left(\lambda_4-\lambda_5\right)^2}{2 (M_\Psi+M_{\psi\psi'})}+\mathcal{O}\left(\lambda_{4,5}^3\right),\label{eq. vertex c2c2h0}\\
  \chi_3\chi_3h^0:\quad&\frac{i v \left(\lambda_4+\lambda_5\right)^2}{2 (M_\Psi-M_{\psi\psi'})}+\mathcal{O}\left(\lambda_{4,5}^3\right),\label{eq. vertex c3c3h0}\\
  \chi_1\chi_1Z_0^\mu:\quad& \left[g_1\sin(\theta_\text{W})+g_2\cos(\theta_\text{W})\right]\frac{i v^2}{4 \left(M_\Psi^2-M_{\psi\psi'}^2\right)}\left(\lambda_4^2-\lambda_5^2\right)\gamma^5\gamma^\mu+\mathcal{O}\left(\lambda_{4,5}^3\right),\label{eq. vertex c1c1Z0}\\
  \chi_2\chi_2Z_0^\mu:\quad& \left[g_1\sin(\theta_\text{W})+g_2\cos(\theta_\text{W})\right]\frac{-i v^2 }{8 M_{\psi\psi'} (M_\Psi+M_{\psi\psi'})}\left(\lambda_4^2-\lambda_5^2\right)\gamma^5\gamma^\mu+\mathcal{O}\left(\lambda_{4,5}^3\right),\label{eq. vertex c2c2Z0}\\
  \chi_3\chi_3Z_0^\mu:\quad& \left[g_1\sin(\theta_\text{W})+g_2\cos(\theta_\text{W})\right]\frac{i v^2 }{8 M_{\psi\psi'} (M_\Psi-M_{\psi\psi'})}\left(\lambda_4^2-\lambda_5^2\right)\gamma^5\gamma^\mu+\mathcal{O}\left(\lambda_{4,5}^3\right).\label{eq. vertex c3c3Z0}
\end{align}
We include the vertices for $\chi_2$ even though $\chi_2$ is never the lightest fermion and thus not abundant. The spin dependent cross section becomes zero if $|\lambda_4|=|\lambda_5|$. If $\chi_3$ is the DM candidate, then the spin independent cross section vanishes for $\lambda_4=-\lambda_5$. For $\chi_1$ the spin independent cross section vanishes for $\lambda_4=\lambda_5\left(\pm\sqrt{\left(\frac{M_{\psi\psi'}}{M_\Psi}\right)-1}-\frac{M_{\psi\psi'}}{M_\Psi}\right)$. We can compare these results with the cross sections calculated by \textsc{SPheno} 4.0.3 \cite{Porod:2003um,Porod:2011nf} and  micrOMEGAs 5.0.8 \cite{Belanger:2018ccd}. 
\begin{figure}
  \centering
  \includegraphics[scale=0.5]{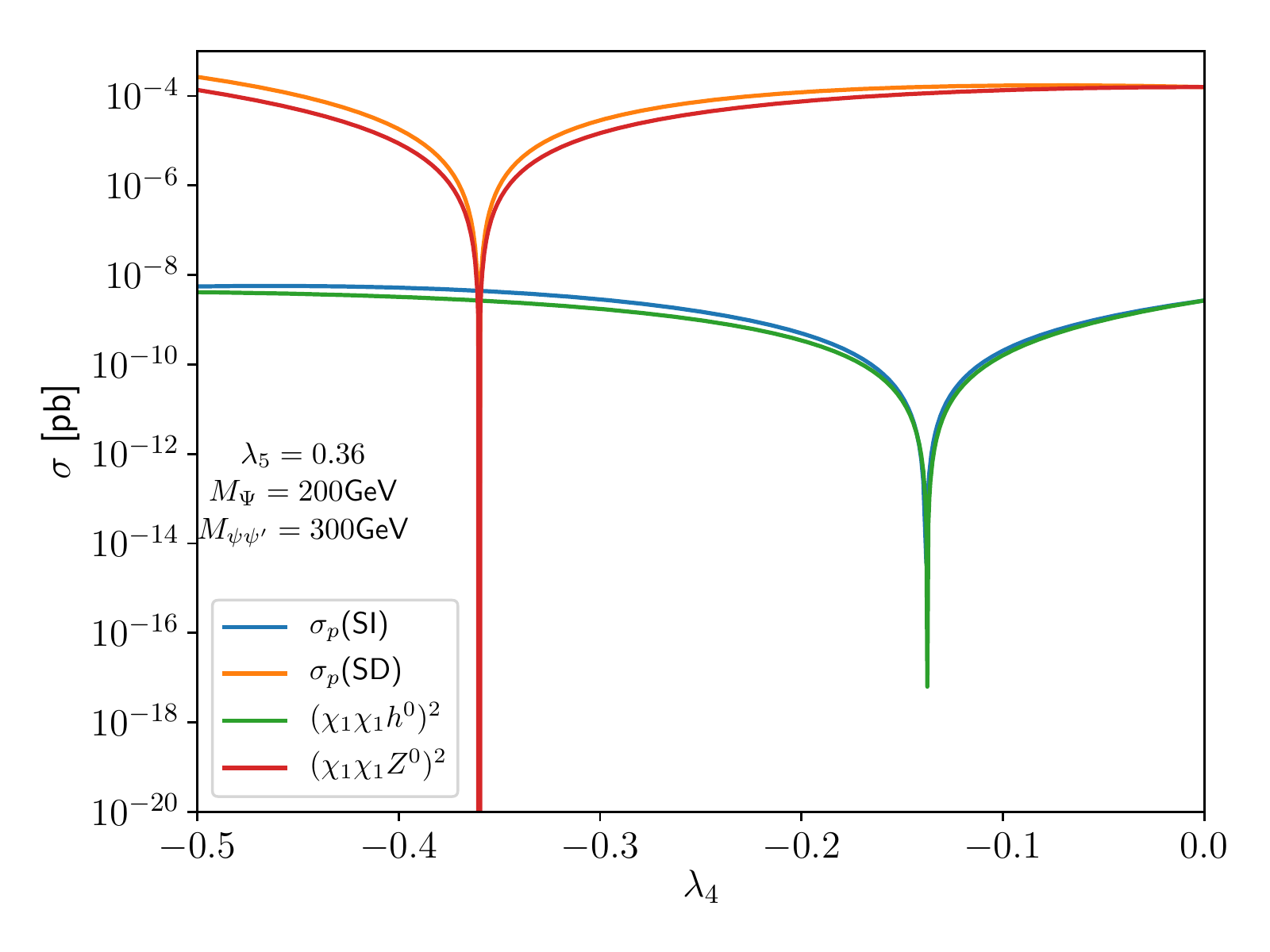}
  \includegraphics[scale=0.5]{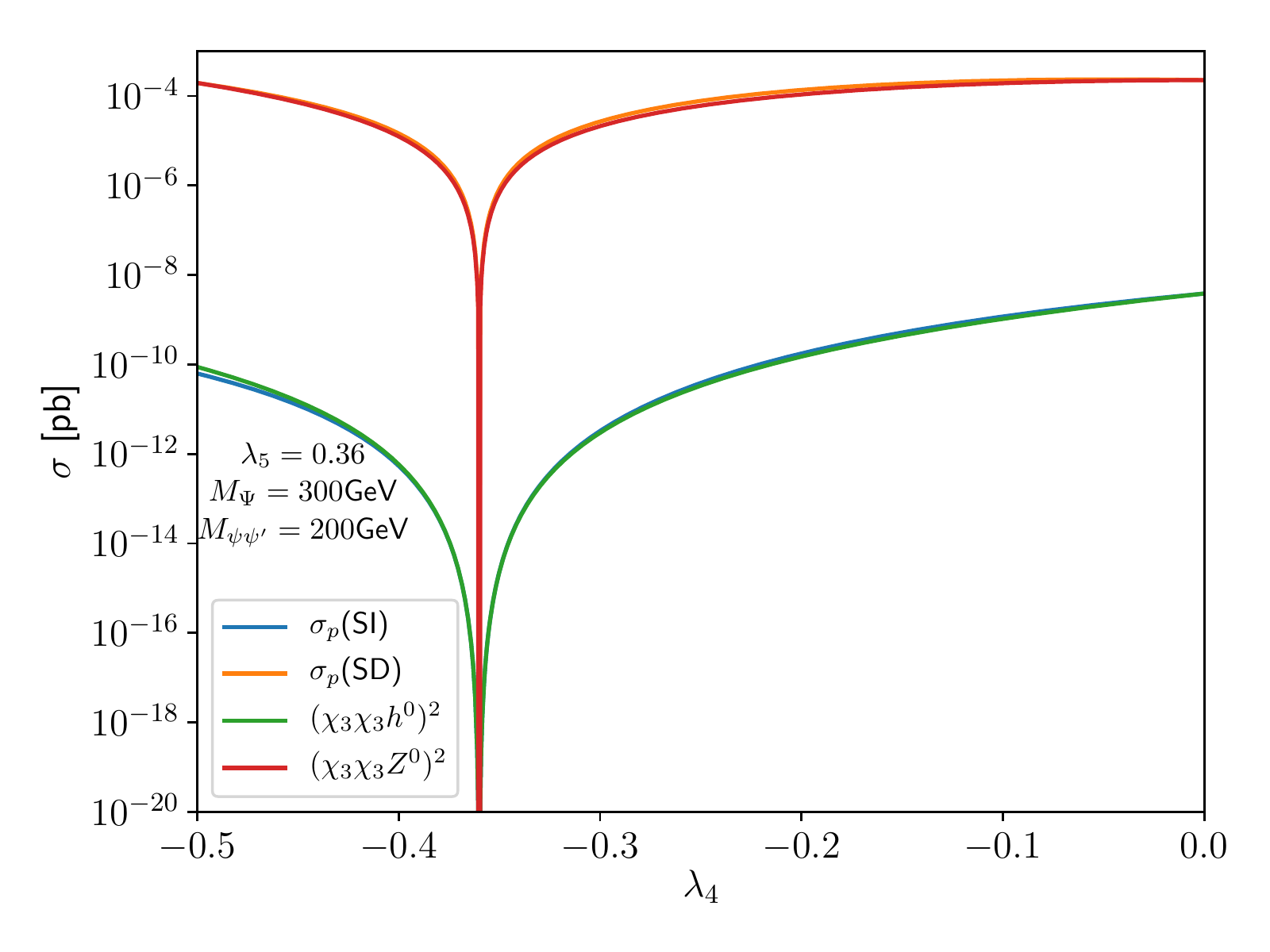}
  \caption{Spin dependent (orange) and independent (blue) cross sections calculated numerically and the vertex factors squared (red and blue) for both $\chi_1$ (left) and $\chi_3$ (right) as DM. The vertex factors have been rescaled so they agree with the cross sections at $\lambda_4=0$. The scalar singlet has been decoupled by setting $m_\phi=10 \,\text{TeV}$ and $\lambda_1$ has been set to zero (cf.\ Ref.\ \cite{May:2020bpo}, Fig.\ 7.1).}
  \label{fig. SISD singlet doublet}
\end{figure}
Fig.\ \ref{fig. SISD singlet doublet} shows the numerical results and our rescaled vertex factors squared. For $\chi_3$ the results agree remarkably well. For $\chi_1$ the qualitative behavior is the same, but we see some deviations, which become larger for larger $|\lambda_4|$. This is not surprising since we expanded the vertex factors for small $\lambda_{4,5}$ and thus our formulas are only correct for $\lambda_{4,5}^2\ll 1$.

\begin{figure}
  \centering
  \includegraphics[scale=0.5]{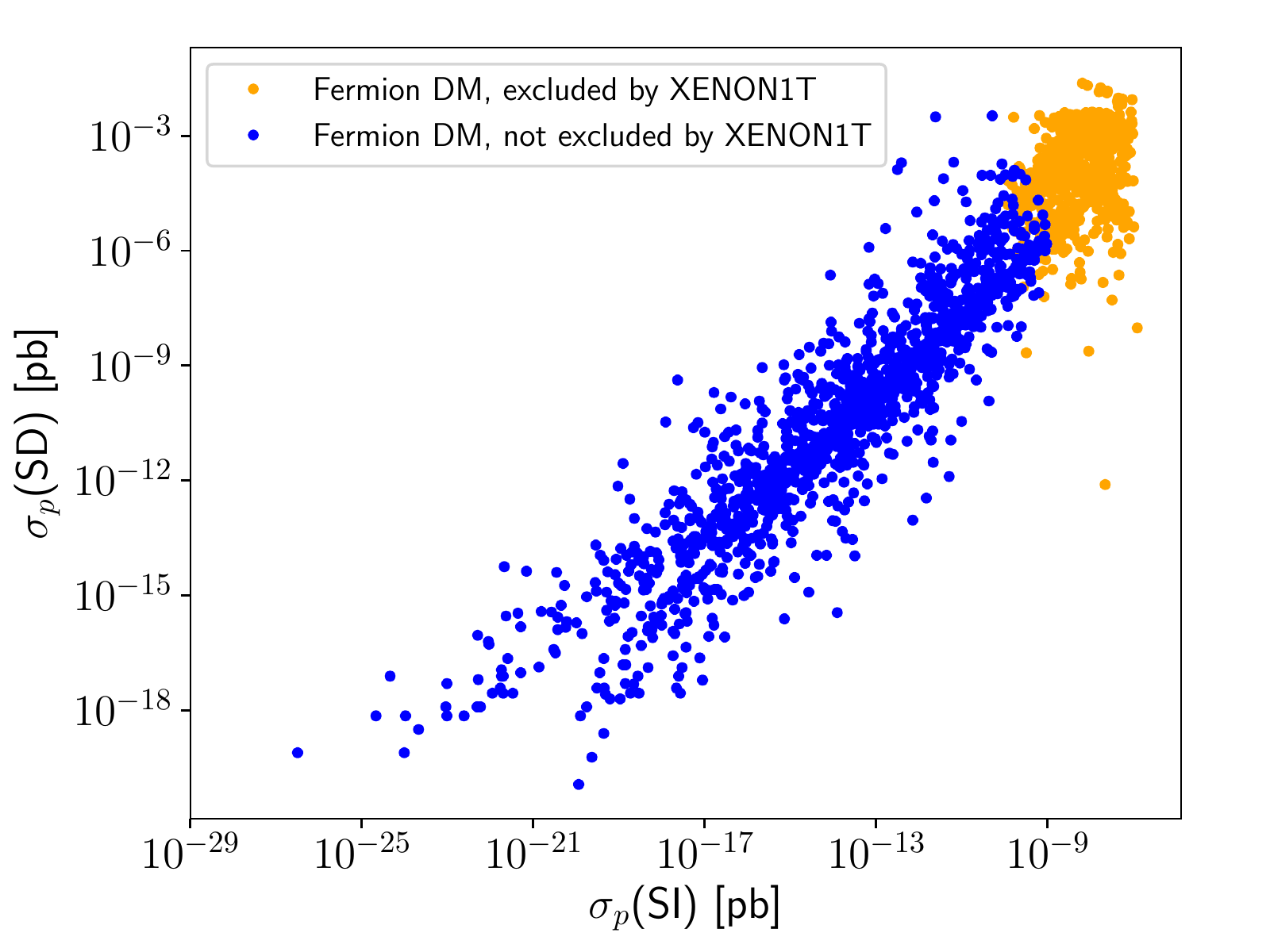}
  \caption{Dependence of the spin dependent cross section on the spin independent cross section using the data from Sec.\ \ref{sec:5}.}
  \label{fig. correlation SDSI}
\end{figure}

From Eqs.\ (\ref{eq. vertex c1c1h0})-(\ref{eq. vertex c3c3Z0}) it is clear that, except from special cases such as $|\lambda_4|=|\lambda_5|$, both cross sections scale with the larger of the two Yukawa couplings $\lambda_{4,5}$. This yields a correlation between the spin dependent and spin independent cross section. This correlation can be made explicit using the data from our scan in Sec.\ \ref{sec:5} as shown in Fig.\ \ref{fig. correlation SDSI}.
This relation explains how the limits on the spin independent cross section set by \textsc{XENON1T} also indirectly restrict the spin dependent cross section.

\bibliography{bib}

\begin{thebibliography}{52}%
\makeatletter
\providecommand \@ifxundefined [1]{%
 \@ifx{#1\undefined}
}%
\providecommand \@ifnum [1]{%
 \ifnum #1\expandafter \@firstoftwo
 \else \expandafter \@secondoftwo
 \fi
}%
\providecommand \@ifx [1]{%
 \ifx #1\expandafter \@firstoftwo
 \else \expandafter \@secondoftwo
 \fi
}%
\providecommand \natexlab [1]{#1}%
\providecommand \enquote  [1]{``#1''}%
\providecommand \bibnamefont  [1]{#1}%
\providecommand \bibfnamefont [1]{#1}%
\providecommand \citenamefont [1]{#1}%
\providecommand \href@noop [0]{\@secondoftwo}%
\providecommand \href [0]{\begingroup \@sanitize@url \@href}%
\providecommand \@href[1]{\@@startlink{#1}\@@href}%
\providecommand \@@href[1]{\endgroup#1\@@endlink}%
\providecommand \@sanitize@url [0]{\catcode `\\12\catcode `\$12\catcode
  `\&12\catcode `\#12\catcode `\^12\catcode `\_12\catcode `\%12\relax}%
\providecommand \@@startlink[1]{}%
\providecommand \@@endlink[0]{}%
\providecommand \url  [0]{\begingroup\@sanitize@url \@url }%
\providecommand \@url [1]{\endgroup\@href {#1}{\urlprefix }}%
\providecommand \urlprefix  [0]{URL }%
\providecommand \Eprint [0]{\href }%
\providecommand \doibase [0]{http://dx.doi.org/}%
\providecommand \selectlanguage [0]{\@gobble}%
\providecommand \bibinfo  [0]{\@secondoftwo}%
\providecommand \bibfield  [0]{\@secondoftwo}%
\providecommand \translation [1]{[#1]}%
\providecommand \BibitemOpen [0]{}%
\providecommand \bibitemStop [0]{}%
\providecommand \bibitemNoStop [0]{.\EOS\space}%
\providecommand \EOS [0]{\spacefactor3000\relax}%
\providecommand \BibitemShut  [1]{\csname bibitem#1\endcsname}%
\let\auto@bib@innerbib\@empty
\bibitem [{\citenamefont {Zyla}\ \emph {et~al.}(2020)\citenamefont {Zyla} \emph
  {et~al.}}]{Zyla:2020zbs}%
  \BibitemOpen
  \bibfield  {author} {\bibinfo {author} {\bibfnamefont {P.~A.}\ \bibnamefont
  {Zyla}} \emph {et~al.} (\bibinfo {collaboration} {Particle Data Group}),\
  }\href {\doibase 10.1093/ptep/ptaa104} {\bibfield  {journal} {\bibinfo
  {journal} {PTEP}\ }\textbf {\bibinfo {volume} {2020}},\ \bibinfo {pages}
  {083C01} (\bibinfo {year} {2020})}\BibitemShut {NoStop}%
\bibitem [{\citenamefont {Klasen}\ \emph {et~al.}(2015)\citenamefont {Klasen},
  \citenamefont {Pohl},\ and\ \citenamefont {Sigl}}]{Klasen:2015uma}%
  \BibitemOpen
  \bibfield  {author} {\bibinfo {author} {\bibfnamefont {M.}~\bibnamefont
  {Klasen}}, \bibinfo {author} {\bibfnamefont {M.}~\bibnamefont {Pohl}}, \ and\
  \bibinfo {author} {\bibfnamefont {G.}~\bibnamefont {Sigl}},\ }\href {\doibase
  10.1016/j.ppnp.2015.07.001} {\bibfield  {journal} {\bibinfo  {journal} {Prog.
  Part. Nucl. Phys.}\ }\textbf {\bibinfo {volume} {85}},\ \bibinfo {pages} {1}
  (\bibinfo {year} {2015})},\ \Eprint {http://arxiv.org/abs/1507.03800}
  {arXiv:1507.03800 [hep-ph]} \BibitemShut {NoStop}%
\bibitem [{\citenamefont {Ma}(2006)}]{Ma:2006km}%
  \BibitemOpen
  \bibfield  {author} {\bibinfo {author} {\bibfnamefont {E.}~\bibnamefont
  {Ma}},\ }\href {\doibase 10.1103/PhysRevD.73.077301} {\bibfield  {journal}
  {\bibinfo  {journal} {Phys. Rev. D}\ }\textbf {\bibinfo {volume} {73}},\
  \bibinfo {pages} {077301} (\bibinfo {year} {2006})},\ \Eprint
  {http://arxiv.org/abs/hep-ph/0601225} {arXiv:hep-ph/0601225} \BibitemShut
  {NoStop}%
\bibitem [{\citenamefont {Klasen}\ \emph {et~al.}(2013)\citenamefont {Klasen},
  \citenamefont {Yaguna}, \citenamefont {Ruiz-Alvarez}, \citenamefont
  {Restrepo},\ and\ \citenamefont {Zapata}}]{Klasen:2013jpa}%
  \BibitemOpen
  \bibfield  {author} {\bibinfo {author} {\bibfnamefont {M.}~\bibnamefont
  {Klasen}}, \bibinfo {author} {\bibfnamefont {C.~E.}\ \bibnamefont {Yaguna}},
  \bibinfo {author} {\bibfnamefont {J.~D.}\ \bibnamefont {Ruiz-Alvarez}},
  \bibinfo {author} {\bibfnamefont {D.}~\bibnamefont {Restrepo}}, \ and\
  \bibinfo {author} {\bibfnamefont {O.}~\bibnamefont {Zapata}},\ }\href
  {\doibase 10.1088/1475-7516/2013/04/044} {\bibfield  {journal} {\bibinfo
  {journal} {JCAP}\ }\textbf {\bibinfo {volume} {04}},\ \bibinfo {pages} {044}
  (\bibinfo {year} {2013})},\ \Eprint {http://arxiv.org/abs/1302.5298}
  {arXiv:1302.5298 [hep-ph]} \BibitemShut {NoStop}%
\bibitem [{\citenamefont {de~Boer}\ \emph {et~al.}(2020)\citenamefont
  {de~Boer}, \citenamefont {Klasen}, \citenamefont {Rodenbeck},\ and\
  \citenamefont {Zeinstra}}]{deBoer:2020yyw}%
  \BibitemOpen
  \bibfield  {author} {\bibinfo {author} {\bibfnamefont {T.}~\bibnamefont
  {de~Boer}}, \bibinfo {author} {\bibfnamefont {M.}~\bibnamefont {Klasen}},
  \bibinfo {author} {\bibfnamefont {C.}~\bibnamefont {Rodenbeck}}, \ and\
  \bibinfo {author} {\bibfnamefont {S.}~\bibnamefont {Zeinstra}},\ }\href
  {\doibase 10.1103/PhysRevD.102.051702} {\bibfield  {journal} {\bibinfo
  {journal} {Phys. Rev. D}\ }\textbf {\bibinfo {volume} {102}},\ \bibinfo
  {pages} {051702} (\bibinfo {year} {2020})},\ \Eprint
  {http://arxiv.org/abs/2007.05338} {arXiv:2007.05338 [hep-ph]} \BibitemShut
  {NoStop}%
\bibitem [{\citenamefont {Restrepo}\ \emph {et~al.}(2013)\citenamefont
  {Restrepo}, \citenamefont {Zapata},\ and\ \citenamefont
  {Yaguna}}]{Restrepo:2013aga}%
  \BibitemOpen
  \bibfield  {author} {\bibinfo {author} {\bibfnamefont {D.}~\bibnamefont
  {Restrepo}}, \bibinfo {author} {\bibfnamefont {O.}~\bibnamefont {Zapata}}, \
  and\ \bibinfo {author} {\bibfnamefont {C.~E.}\ \bibnamefont {Yaguna}},\
  }\href {\doibase 10.1007/JHEP11(2013)011} {\bibfield  {journal} {\bibinfo
  {journal} {JHEP}\ }\textbf {\bibinfo {volume} {11}},\ \bibinfo {pages} {011}
  (\bibinfo {year} {2013})},\ \Eprint {http://arxiv.org/abs/1308.3655}
  {arXiv:1308.3655 [hep-ph]} \BibitemShut {NoStop}%
\bibitem [{\citenamefont {Arina}\ \emph {et~al.}(2015)\citenamefont {Arina},
  \citenamefont {Kulkarni},\ and\ \citenamefont {Silk}}]{Arina:2015zoa}%
  \BibitemOpen
  \bibfield  {author} {\bibinfo {author} {\bibfnamefont {C.}~\bibnamefont
  {Arina}}, \bibinfo {author} {\bibfnamefont {S.}~\bibnamefont {Kulkarni}}, \
  and\ \bibinfo {author} {\bibfnamefont {J.}~\bibnamefont {Silk}},\ }\href
  {\doibase 10.1103/PhysRevD.92.083519} {\bibfield  {journal} {\bibinfo
  {journal} {Phys. Rev. D}\ }\textbf {\bibinfo {volume} {92}},\ \bibinfo
  {pages} {083519} (\bibinfo {year} {2015})},\ \Eprint
  {http://arxiv.org/abs/1506.08202} {arXiv:1506.08202 [astro-ph.HE]}
  \BibitemShut {NoStop}%
\bibitem [{\citenamefont {Lindner}\ \emph {et~al.}(2010)\citenamefont
  {Lindner}, \citenamefont {Merle},\ and\ \citenamefont
  {Niro}}]{Lindner:2010rr}%
  \BibitemOpen
  \bibfield  {author} {\bibinfo {author} {\bibfnamefont {M.}~\bibnamefont
  {Lindner}}, \bibinfo {author} {\bibfnamefont {A.}~\bibnamefont {Merle}}, \
  and\ \bibinfo {author} {\bibfnamefont {V.}~\bibnamefont {Niro}},\ }\href
  {\doibase 10.1103/PhysRevD.82.123529} {\bibfield  {journal} {\bibinfo
  {journal} {Phys. Rev. D}\ }\textbf {\bibinfo {volume} {82}},\ \bibinfo
  {pages} {123529} (\bibinfo {year} {2010})},\ \Eprint
  {http://arxiv.org/abs/1005.3116} {arXiv:1005.3116 [hep-ph]} \BibitemShut
  {NoStop}%
\bibitem [{\citenamefont {El~Aisati}\ \emph {et~al.}(2017)\citenamefont
  {El~Aisati}, \citenamefont {Garcia-Cely}, \citenamefont {Hambye},\ and\
  \citenamefont {Vanderheyden}}]{ElAisati:2017ppn}%
  \BibitemOpen
  \bibfield  {author} {\bibinfo {author} {\bibfnamefont {C.}~\bibnamefont
  {El~Aisati}}, \bibinfo {author} {\bibfnamefont {C.}~\bibnamefont
  {Garcia-Cely}}, \bibinfo {author} {\bibfnamefont {T.}~\bibnamefont {Hambye}},
  \ and\ \bibinfo {author} {\bibfnamefont {L.}~\bibnamefont {Vanderheyden}},\
  }\href {\doibase 10.1088/1475-7516/2017/10/021} {\bibfield  {journal}
  {\bibinfo  {journal} {JCAP}\ }\textbf {\bibinfo {volume} {10}},\ \bibinfo
  {pages} {021} (\bibinfo {year} {2017})},\ \Eprint
  {http://arxiv.org/abs/1706.06600} {arXiv:1706.06600 [hep-ph]} \BibitemShut
  {NoStop}%
\bibitem [{\citenamefont {Farzan}(2012)}]{Farzan:2011ck}%
  \BibitemOpen
  \bibfield  {author} {\bibinfo {author} {\bibfnamefont {Y.}~\bibnamefont
  {Farzan}},\ }\href {\doibase 10.1007/JHEP02(2012)091} {\bibfield  {journal}
  {\bibinfo  {journal} {JHEP}\ }\textbf {\bibinfo {volume} {02}},\ \bibinfo
  {pages} {091} (\bibinfo {year} {2012})},\ \Eprint
  {http://arxiv.org/abs/1111.1063} {arXiv:1111.1063 [hep-ph]} \BibitemShut
  {NoStop}%
\bibitem [{\citenamefont {Andreas}\ \emph {et~al.}(2009)\citenamefont
  {Andreas}, \citenamefont {Tytgat},\ and\ \citenamefont
  {Swillens}}]{Andreas:2009hj}%
  \BibitemOpen
  \bibfield  {author} {\bibinfo {author} {\bibfnamefont {S.}~\bibnamefont
  {Andreas}}, \bibinfo {author} {\bibfnamefont {M.~H.}\ \bibnamefont {Tytgat}},
  \ and\ \bibinfo {author} {\bibfnamefont {Q.}~\bibnamefont {Swillens}},\
  }\href {\doibase 10.1088/1475-7516/2009/04/004} {\bibfield  {journal}
  {\bibinfo  {journal} {JCAP}\ }\textbf {\bibinfo {volume} {04}},\ \bibinfo
  {pages} {004} (\bibinfo {year} {2009})},\ \Eprint
  {http://arxiv.org/abs/0901.1750} {arXiv:0901.1750 [hep-ph]} \BibitemShut
  {NoStop}%
\bibitem [{\citenamefont {Albert}\ \emph
  {et~al.}(2017{\natexlab{a}})\citenamefont {Albert} \emph
  {et~al.}}]{Albert:2016dsy}%
  \BibitemOpen
  \bibfield  {author} {\bibinfo {author} {\bibfnamefont {A.}~\bibnamefont
  {Albert}} \emph {et~al.} (\bibinfo {collaboration} {ANTARES}),\ }\href
  {\doibase 10.1016/j.dark.2017.04.005} {\bibfield  {journal} {\bibinfo
  {journal} {Phys. Dark Univ.}\ }\textbf {\bibinfo {volume} {16}},\ \bibinfo
  {pages} {41} (\bibinfo {year} {2017}{\natexlab{a}})},\ \Eprint
  {http://arxiv.org/abs/1612.06792} {arXiv:1612.06792 [hep-ex]} \BibitemShut
  {NoStop}%
\bibitem [{\citenamefont {Aartsen}\ \emph
  {et~al.}(2017{\natexlab{a}})\citenamefont {Aartsen} \emph
  {et~al.}}]{Aartsen:2017ulx}%
  \BibitemOpen
  \bibfield  {author} {\bibinfo {author} {\bibfnamefont {M.~G.}\ \bibnamefont
  {Aartsen}} \emph {et~al.} (\bibinfo {collaboration} {IceCube}),\ }\href
  {\doibase 10.1140/epjc/s10052-017-5213-y} {\bibfield  {journal} {\bibinfo
  {journal} {Eur. Phys. J. C}\ }\textbf {\bibinfo {volume} {77}},\ \bibinfo
  {pages} {627} (\bibinfo {year} {2017}{\natexlab{a}})},\ \Eprint
  {http://arxiv.org/abs/1705.08103} {arXiv:1705.08103 [hep-ex]} \BibitemShut
  {NoStop}%
\bibitem [{\citenamefont {Albert}\ \emph
  {et~al.}(2017{\natexlab{b}})\citenamefont {Albert} \emph
  {et~al.}}]{Albert:2016emp}%
  \BibitemOpen
  \bibfield  {author} {\bibinfo {author} {\bibfnamefont {A.}~\bibnamefont
  {Albert}} \emph {et~al.},\ }\href {\doibase 10.1016/j.physletb.2017.03.063}
  {\bibfield  {journal} {\bibinfo  {journal} {Phys. Lett. B}\ }\textbf
  {\bibinfo {volume} {769}},\ \bibinfo {pages} {249} (\bibinfo {year}
  {2017}{\natexlab{b}})},\ \bibinfo {note} {[Erratum: Phys.Lett.B 796, 253--255
  (2019)]},\ \Eprint {http://arxiv.org/abs/1612.04595} {arXiv:1612.04595
  [astro-ph.HE]} \BibitemShut {NoStop}%
\bibitem [{\citenamefont {Albert}\ \emph {et~al.}(2020)\citenamefont {Albert}
  \emph {et~al.}}]{Aartsen:2020tdl}%
  \BibitemOpen
  \bibfield  {author} {\bibinfo {author} {\bibfnamefont {A.}~\bibnamefont
  {Albert}} \emph {et~al.} (\bibinfo {collaboration} {ANTARES, IceCube}),\
  }\href {\doibase 10.1103/PhysRevD.102.082002} {\bibfield  {journal} {\bibinfo
   {journal} {Phys. Rev. D}\ }\textbf {\bibinfo {volume} {102}},\ \bibinfo
  {pages} {082002} (\bibinfo {year} {2020})},\ \Eprint
  {http://arxiv.org/abs/2003.06614} {arXiv:2003.06614 [astro-ph.HE]}
  \BibitemShut {NoStop}%
\bibitem [{\citenamefont {Abe}\ \emph {et~al.}(2020)\citenamefont {Abe} \emph
  {et~al.}}]{Abe:2020sbr}%
  \BibitemOpen
  \bibfield  {author} {\bibinfo {author} {\bibfnamefont {K.}~\bibnamefont
  {Abe}} \emph {et~al.} (\bibinfo {collaboration} {Super-Kamiokande}),\ }\href
  {\doibase 10.1103/PhysRevD.102.072002} {\bibfield  {journal} {\bibinfo
  {journal} {Phys. Rev. D}\ }\textbf {\bibinfo {volume} {102}},\ \bibinfo
  {pages} {072002} (\bibinfo {year} {2020})},\ \Eprint
  {http://arxiv.org/abs/2005.05109} {arXiv:2005.05109 [hep-ex]} \BibitemShut
  {NoStop}%
\bibitem [{\citenamefont {{Gould}}(1987)}]{Gould:1987}%
  \BibitemOpen
  \bibfield  {author} {\bibinfo {author} {\bibfnamefont {A.}~\bibnamefont
  {{Gould}}},\ }\href {\doibase 10.1086/165653} {\bibfield  {journal} {\bibinfo
   {journal} {ApJ}\ }\textbf {\bibinfo {volume} {321}},\ \bibinfo {pages} {571}
  (\bibinfo {year} {1987})}\BibitemShut {NoStop}%
\bibitem [{\citenamefont {Gould}(1992)}]{Gould:1991hx}%
  \BibitemOpen
  \bibfield  {author} {\bibinfo {author} {\bibfnamefont {A.}~\bibnamefont
  {Gould}},\ }\href {\doibase 10.1086/171156} {\bibfield  {journal} {\bibinfo
  {journal} {Astrophys. J.}\ }\textbf {\bibinfo {volume} {388}},\ \bibinfo
  {pages} {338} (\bibinfo {year} {1992})}\BibitemShut {NoStop}%
\bibitem [{\citenamefont {Aprile}\ \emph {et~al.}(2018)\citenamefont {Aprile}
  \emph {et~al.}}]{Aprile:2018dbl}%
  \BibitemOpen
  \bibfield  {author} {\bibinfo {author} {\bibfnamefont {E.}~\bibnamefont
  {Aprile}} \emph {et~al.} (\bibinfo {collaboration} {XENON}),\ }\href
  {\doibase 10.1103/PhysRevLett.121.111302} {\bibfield  {journal} {\bibinfo
  {journal} {Phys. Rev. Lett.}\ }\textbf {\bibinfo {volume} {121}},\ \bibinfo
  {pages} {111302} (\bibinfo {year} {2018})},\ \Eprint
  {http://arxiv.org/abs/1805.12562} {arXiv:1805.12562 [astro-ph.CO]}
  \BibitemShut {NoStop}%
\bibitem [{\citenamefont {Aprile}\ \emph
  {et~al.}(2019{\natexlab{a}})\citenamefont {Aprile} \emph
  {et~al.}}]{Aprile:2019dbj}%
  \BibitemOpen
  \bibfield  {author} {\bibinfo {author} {\bibfnamefont {E.}~\bibnamefont
  {Aprile}} \emph {et~al.} (\bibinfo {collaboration} {XENON}),\ }\href
  {\doibase 10.1103/PhysRevLett.122.141301} {\bibfield  {journal} {\bibinfo
  {journal} {Phys. Rev. Lett.}\ }\textbf {\bibinfo {volume} {122}},\ \bibinfo
  {pages} {141301} (\bibinfo {year} {2019}{\natexlab{a}})},\ \Eprint
  {http://arxiv.org/abs/1902.03234} {arXiv:1902.03234 [astro-ph.CO]}
  \BibitemShut {NoStop}%
\bibitem [{\citenamefont {Aprile}\ \emph
  {et~al.}(2019{\natexlab{b}})\citenamefont {Aprile} \emph
  {et~al.}}]{Aprile:2019xxb}%
  \BibitemOpen
  \bibfield  {author} {\bibinfo {author} {\bibfnamefont {E.}~\bibnamefont
  {Aprile}} \emph {et~al.} (\bibinfo {collaboration} {XENON}),\ }\href
  {\doibase 10.1103/PhysRevLett.123.251801} {\bibfield  {journal} {\bibinfo
  {journal} {Phys. Rev. Lett.}\ }\textbf {\bibinfo {volume} {123}},\ \bibinfo
  {pages} {251801} (\bibinfo {year} {2019}{\natexlab{b}})},\ \Eprint
  {http://arxiv.org/abs/1907.11485} {arXiv:1907.11485 [hep-ex]} \BibitemShut
  {NoStop}%
\bibitem [{\citenamefont {Amole}\ \emph {et~al.}(2019)\citenamefont {Amole}
  \emph {et~al.}}]{Amole:2019fdf}%
  \BibitemOpen
  \bibfield  {author} {\bibinfo {author} {\bibfnamefont {C.}~\bibnamefont
  {Amole}} \emph {et~al.} (\bibinfo {collaboration} {PICO}),\ }\href {\doibase
  10.1103/PhysRevD.100.022001} {\bibfield  {journal} {\bibinfo  {journal}
  {Phys. Rev. D}\ }\textbf {\bibinfo {volume} {100}},\ \bibinfo {pages}
  {022001} (\bibinfo {year} {2019})},\ \Eprint
  {http://arxiv.org/abs/1902.04031} {arXiv:1902.04031 [astro-ph.CO]}
  \BibitemShut {NoStop}%
\bibitem [{\citenamefont {Adrian-Martinez}\ \emph {et~al.}(2016)\citenamefont
  {Adrian-Martinez} \emph {et~al.}}]{Adrian-Martinez:2016gti}%
  \BibitemOpen
  \bibfield  {author} {\bibinfo {author} {\bibfnamefont {S.}~\bibnamefont
  {Adrian-Martinez}} \emph {et~al.} (\bibinfo {collaboration} {ANTARES}),\
  }\href {\doibase 10.1016/j.physletb.2016.05.019} {\bibfield  {journal}
  {\bibinfo  {journal} {Phys. Lett. B}\ }\textbf {\bibinfo {volume} {759}},\
  \bibinfo {pages} {69} (\bibinfo {year} {2016})},\ \Eprint
  {http://arxiv.org/abs/1603.02228} {arXiv:1603.02228 [astro-ph.HE]}
  \BibitemShut {NoStop}%
\bibitem [{\citenamefont {Aartsen}\ \emph
  {et~al.}(2017{\natexlab{b}})\citenamefont {Aartsen} \emph
  {et~al.}}]{Aartsen:2016zhm}%
  \BibitemOpen
  \bibfield  {author} {\bibinfo {author} {\bibfnamefont {M.}~\bibnamefont
  {Aartsen}} \emph {et~al.} (\bibinfo {collaboration} {IceCube}),\ }\href
  {\doibase 10.1140/epjc/s10052-017-4689-9} {\bibfield  {journal} {\bibinfo
  {journal} {Eur. Phys. J. C}\ }\textbf {\bibinfo {volume} {77}},\ \bibinfo
  {pages} {146} (\bibinfo {year} {2017}{\natexlab{b}})},\ \bibinfo {note}
  {[Erratum: Eur.Phys.J.C 79, 214 (2019)]},\ \Eprint
  {http://arxiv.org/abs/1612.05949} {arXiv:1612.05949 [astro-ph.HE]}
  \BibitemShut {NoStop}%
\bibitem [{\citenamefont {Choi}\ \emph {et~al.}(2015)\citenamefont {Choi} \emph
  {et~al.}}]{Choi:2015ara}%
  \BibitemOpen
  \bibfield  {author} {\bibinfo {author} {\bibfnamefont {K.}~\bibnamefont
  {Choi}} \emph {et~al.} (\bibinfo {collaboration} {Super-Kamiokande}),\ }\href
  {\doibase 10.1103/PhysRevLett.114.141301} {\bibfield  {journal} {\bibinfo
  {journal} {Phys. Rev. Lett.}\ }\textbf {\bibinfo {volume} {114}},\ \bibinfo
  {pages} {141301} (\bibinfo {year} {2015})},\ \Eprint
  {http://arxiv.org/abs/1503.04858} {arXiv:1503.04858 [hep-ex]} \BibitemShut
  {NoStop}%
\bibitem [{\citenamefont {Fiaschi}\ \emph {et~al.}(2019)\citenamefont
  {Fiaschi}, \citenamefont {Klasen},\ and\ \citenamefont
  {May}}]{Fiaschi:2018rky}%
  \BibitemOpen
  \bibfield  {author} {\bibinfo {author} {\bibfnamefont {J.}~\bibnamefont
  {Fiaschi}}, \bibinfo {author} {\bibfnamefont {M.}~\bibnamefont {Klasen}}, \
  and\ \bibinfo {author} {\bibfnamefont {S.}~\bibnamefont {May}},\ }\href
  {\doibase 10.1007/JHEP05(2019)015} {\bibfield  {journal} {\bibinfo  {journal}
  {JHEP}\ }\textbf {\bibinfo {volume} {05}},\ \bibinfo {pages} {015} (\bibinfo
  {year} {2019})},\ \Eprint {http://arxiv.org/abs/1812.11133} {arXiv:1812.11133
  [hep-ph]} \BibitemShut {NoStop}%
\bibitem [{\citenamefont {May}(2020)}]{May:2020bpo}%
  \BibitemOpen
  \bibfield  {author} {\bibinfo {author} {\bibfnamefont {S.}~\bibnamefont
  {May}},\ }\emph {\bibinfo {title} {{Minimal dark matter models with radiative
  neutrino masses: From Lagrangians to observables}}},\ \href@noop {} {\bibinfo
  {type} {Master thesis}} (\bibinfo {year} {2020}),\ \Eprint
  {http://arxiv.org/abs/2003.04157} {arXiv:2003.04157 [hep-ph]} \BibitemShut
  {NoStop}%
\bibitem [{\citenamefont {Suematsu}\ \emph {et~al.}(2009)\citenamefont
  {Suematsu}, \citenamefont {Toma},\ and\ \citenamefont
  {Yoshida}}]{Suematsu:2009ww}%
  \BibitemOpen
  \bibfield  {author} {\bibinfo {author} {\bibfnamefont {D.}~\bibnamefont
  {Suematsu}}, \bibinfo {author} {\bibfnamefont {T.}~\bibnamefont {Toma}}, \
  and\ \bibinfo {author} {\bibfnamefont {T.}~\bibnamefont {Yoshida}},\ }\href
  {\doibase 10.1103/PhysRevD.79.093004} {\bibfield  {journal} {\bibinfo
  {journal} {Phys. Rev. D}\ }\textbf {\bibinfo {volume} {79}},\ \bibinfo
  {pages} {093004} (\bibinfo {year} {2009})},\ \Eprint
  {http://arxiv.org/abs/0903.0287} {arXiv:0903.0287 [hep-ph]} \BibitemShut
  {NoStop}%
\bibitem [{\citenamefont {Harz}\ \emph {et~al.}(2013)\citenamefont {Harz},
  \citenamefont {Herrmann}, \citenamefont {Klasen}, \citenamefont {Kovarik},\
  and\ \citenamefont {Boulc'h}}]{Harz:2012fz}%
  \BibitemOpen
  \bibfield  {author} {\bibinfo {author} {\bibfnamefont {J.}~\bibnamefont
  {Harz}}, \bibinfo {author} {\bibfnamefont {B.}~\bibnamefont {Herrmann}},
  \bibinfo {author} {\bibfnamefont {M.}~\bibnamefont {Klasen}}, \bibinfo
  {author} {\bibfnamefont {K.}~\bibnamefont {Kovarik}}, \ and\ \bibinfo
  {author} {\bibfnamefont {Q.~L.}\ \bibnamefont {Boulc'h}},\ }\href {\doibase
  10.1103/PhysRevD.87.054031} {\bibfield  {journal} {\bibinfo  {journal} {Phys.
  Rev. D}\ }\textbf {\bibinfo {volume} {87}},\ \bibinfo {pages} {054031}
  (\bibinfo {year} {2013})},\ \Eprint {http://arxiv.org/abs/1212.5241}
  {arXiv:1212.5241 [hep-ph]} \BibitemShut {NoStop}%
\bibitem [{\citenamefont {Herrmann}\ \emph {et~al.}(2014)\citenamefont
  {Herrmann}, \citenamefont {Klasen}, \citenamefont {Kovarik}, \citenamefont
  {Meinecke},\ and\ \citenamefont {Steppeler}}]{Herrmann:2014kma}%
  \BibitemOpen
  \bibfield  {author} {\bibinfo {author} {\bibfnamefont {B.}~\bibnamefont
  {Herrmann}}, \bibinfo {author} {\bibfnamefont {M.}~\bibnamefont {Klasen}},
  \bibinfo {author} {\bibfnamefont {K.}~\bibnamefont {Kovarik}}, \bibinfo
  {author} {\bibfnamefont {M.}~\bibnamefont {Meinecke}}, \ and\ \bibinfo
  {author} {\bibfnamefont {P.}~\bibnamefont {Steppeler}},\ }\href {\doibase
  10.1103/PhysRevD.89.114012} {\bibfield  {journal} {\bibinfo  {journal} {Phys.
  Rev. D}\ }\textbf {\bibinfo {volume} {89}},\ \bibinfo {pages} {114012}
  (\bibinfo {year} {2014})},\ \Eprint {http://arxiv.org/abs/1404.2931}
  {arXiv:1404.2931 [hep-ph]} \BibitemShut {NoStop}%
\bibitem [{\citenamefont {Harz}\ \emph {et~al.}(2015)\citenamefont {Harz},
  \citenamefont {Herrmann}, \citenamefont {Klasen},\ and\ \citenamefont
  {Kovarik}}]{Harz:2014tma}%
  \BibitemOpen
  \bibfield  {author} {\bibinfo {author} {\bibfnamefont {J.}~\bibnamefont
  {Harz}}, \bibinfo {author} {\bibfnamefont {B.}~\bibnamefont {Herrmann}},
  \bibinfo {author} {\bibfnamefont {M.}~\bibnamefont {Klasen}}, \ and\ \bibinfo
  {author} {\bibfnamefont {K.}~\bibnamefont {Kovarik}},\ }\href {\doibase
  10.1103/PhysRevD.91.034028} {\bibfield  {journal} {\bibinfo  {journal} {Phys.
  Rev. D}\ }\textbf {\bibinfo {volume} {91}},\ \bibinfo {pages} {034028}
  (\bibinfo {year} {2015})},\ \Eprint {http://arxiv.org/abs/1409.2898}
  {arXiv:1409.2898 [hep-ph]} \BibitemShut {NoStop}%
\bibitem [{\citenamefont {Branahl}\ \emph {et~al.}(2019)\citenamefont
  {Branahl}, \citenamefont {Harz}, \citenamefont {Herrmann}, \citenamefont
  {Klasen}, \citenamefont {Kova\v{r}\'\i{}k},\ and\ \citenamefont
  {Schmiemann}}]{Branahl:2019yot}%
  \BibitemOpen
  \bibfield  {author} {\bibinfo {author} {\bibfnamefont {J.}~\bibnamefont
  {Branahl}}, \bibinfo {author} {\bibfnamefont {J.}~\bibnamefont {Harz}},
  \bibinfo {author} {\bibfnamefont {B.}~\bibnamefont {Herrmann}}, \bibinfo
  {author} {\bibfnamefont {M.}~\bibnamefont {Klasen}}, \bibinfo {author}
  {\bibfnamefont {K.}~\bibnamefont {Kova\v{r}\'\i{}k}}, \ and\ \bibinfo
  {author} {\bibfnamefont {S.}~\bibnamefont {Schmiemann}},\ }\href {\doibase
  10.1103/PhysRevD.100.115003} {\bibfield  {journal} {\bibinfo  {journal}
  {Phys. Rev. D}\ }\textbf {\bibinfo {volume} {100}},\ \bibinfo {pages}
  {115003} (\bibinfo {year} {2019})},\ \Eprint
  {http://arxiv.org/abs/1909.09527} {arXiv:1909.09527 [hep-ph]} \BibitemShut
  {NoStop}%
\bibitem [{\citenamefont {Cirelli}\ \emph {et~al.}(2006)\citenamefont
  {Cirelli}, \citenamefont {Fornengo},\ and\ \citenamefont
  {Strumia}}]{Cirelli:2005uq}%
  \BibitemOpen
  \bibfield  {author} {\bibinfo {author} {\bibfnamefont {M.}~\bibnamefont
  {Cirelli}}, \bibinfo {author} {\bibfnamefont {N.}~\bibnamefont {Fornengo}}, \
  and\ \bibinfo {author} {\bibfnamefont {A.}~\bibnamefont {Strumia}},\ }\href
  {\doibase 10.1016/j.nuclphysb.2006.07.012} {\bibfield  {journal} {\bibinfo
  {journal} {Nucl. Phys. B}\ }\textbf {\bibinfo {volume} {753}},\ \bibinfo
  {pages} {178} (\bibinfo {year} {2006})},\ \Eprint
  {http://arxiv.org/abs/hep-ph/0512090} {arXiv:hep-ph/0512090} \BibitemShut
  {NoStop}%
\bibitem [{\citenamefont {Casas}\ and\ \citenamefont
  {Ibarra}(2001)}]{Casas:2001sr}%
  \BibitemOpen
  \bibfield  {author} {\bibinfo {author} {\bibfnamefont {J.}~\bibnamefont
  {Casas}}\ and\ \bibinfo {author} {\bibfnamefont {A.}~\bibnamefont {Ibarra}},\
  }\href {\doibase 10.1016/S0550-3213(01)00475-8} {\bibfield  {journal}
  {\bibinfo  {journal} {Nucl. Phys. B}\ }\textbf {\bibinfo {volume} {618}},\
  \bibinfo {pages} {171} (\bibinfo {year} {2001})},\ \Eprint
  {http://arxiv.org/abs/hep-ph/0103065} {arXiv:hep-ph/0103065} \BibitemShut
  {NoStop}%
\bibitem [{\citenamefont {Jungman}\ \emph {et~al.}(1996)\citenamefont
  {Jungman}, \citenamefont {Kamionkowski},\ and\ \citenamefont
  {Griest}}]{JUNGMAN1996195}%
  \BibitemOpen
  \bibfield  {author} {\bibinfo {author} {\bibfnamefont {G.}~\bibnamefont
  {Jungman}}, \bibinfo {author} {\bibfnamefont {M.}~\bibnamefont
  {Kamionkowski}}, \ and\ \bibinfo {author} {\bibfnamefont {K.}~\bibnamefont
  {Griest}},\ }\href {\doibase https://doi.org/10.1016/0370-1573(95)00058-5}
  {\bibfield  {journal} {\bibinfo  {journal} {Physics Reports}\ }\textbf
  {\bibinfo {volume} {267}},\ \bibinfo {pages} {195 } (\bibinfo {year}
  {1996})}\BibitemShut {NoStop}%
\bibitem [{\citenamefont {B\'elanger}\ \emph {et~al.}(2018)\citenamefont
  {B\'elanger}, \citenamefont {Boudjema}, \citenamefont {Goudelis},
  \citenamefont {Pukhov},\ and\ \citenamefont {Zaldivar}}]{Belanger:2018ccd}%
  \BibitemOpen
  \bibfield  {author} {\bibinfo {author} {\bibfnamefont {G.}~\bibnamefont
  {B\'elanger}}, \bibinfo {author} {\bibfnamefont {F.}~\bibnamefont
  {Boudjema}}, \bibinfo {author} {\bibfnamefont {A.}~\bibnamefont {Goudelis}},
  \bibinfo {author} {\bibfnamefont {A.}~\bibnamefont {Pukhov}}, \ and\ \bibinfo
  {author} {\bibfnamefont {B.}~\bibnamefont {Zaldivar}},\ }\href {\doibase
  10.1016/j.cpc.2018.04.027} {\bibfield  {journal} {\bibinfo  {journal}
  {Comput. Phys. Commun.}\ }\textbf {\bibinfo {volume} {231}},\ \bibinfo
  {pages} {173} (\bibinfo {year} {2018})},\ \Eprint
  {http://arxiv.org/abs/1801.03509} {arXiv:1801.03509 [hep-ph]} \BibitemShut
  {NoStop}%
\bibitem [{\citenamefont {{Bugaev, Edgar and Montaruli, Teresa and Shlepin,
  Yuri and Sokalski, Igor A.}}(2004)}]{Buga2003}%
  \BibitemOpen
  \bibfield  {author} {\bibinfo {author} {\bibnamefont {{Bugaev, Edgar and
  Montaruli, Teresa and Shlepin, Yuri and Sokalski, Igor A.}}},\ }\href
  {\doibase 10.1016/j.astropartphys.2004.03.002} {\bibfield  {journal}
  {\bibinfo  {journal} {Astropart. Phys.}\ }\textbf {\bibinfo {volume} {21}},\
  \bibinfo {pages} {491} (\bibinfo {year} {2004})},\ \Eprint
  {http://arxiv.org/abs/hep-ph/0312295} {arXiv:hep-ph/0312295} \BibitemShut
  {NoStop}%
\bibitem [{\citenamefont {Cirelli}\ \emph {et~al.}(2005)\citenamefont
  {Cirelli}, \citenamefont {Fornengo}, \citenamefont {Montaruli}, \citenamefont
  {Sokalski}, \citenamefont {Strumia},\ and\ \citenamefont
  {Vissani}}]{Cirelli:2005gh}%
  \BibitemOpen
  \bibfield  {author} {\bibinfo {author} {\bibfnamefont {M.}~\bibnamefont
  {Cirelli}}, \bibinfo {author} {\bibfnamefont {N.}~\bibnamefont {Fornengo}},
  \bibinfo {author} {\bibfnamefont {T.}~\bibnamefont {Montaruli}}, \bibinfo
  {author} {\bibfnamefont {I.~A.}\ \bibnamefont {Sokalski}}, \bibinfo {author}
  {\bibfnamefont {A.}~\bibnamefont {Strumia}}, \ and\ \bibinfo {author}
  {\bibfnamefont {F.}~\bibnamefont {Vissani}},\ }\href {\doibase
  10.1016/j.nuclphysb.2007.10.001} {\bibfield  {journal} {\bibinfo  {journal}
  {Nucl. Phys. B}\ }\textbf {\bibinfo {volume} {727}},\ \bibinfo {pages} {99}
  (\bibinfo {year} {2005})},\ \bibinfo {note} {[Erratum: Nucl.Phys.B 790,
  338--344 (2008)]},\ \Eprint {http://arxiv.org/abs/hep-ph/0506298}
  {arXiv:hep-ph/0506298} \BibitemShut {NoStop}%
\bibitem [{\citenamefont {Porod}(2003)}]{Porod:2003um}%
  \BibitemOpen
  \bibfield  {author} {\bibinfo {author} {\bibfnamefont {W.}~\bibnamefont
  {Porod}},\ }\href {\doibase 10.1016/S0010-4655(03)00222-4} {\bibfield
  {journal} {\bibinfo  {journal} {Comput. Phys. Commun.}\ }\textbf {\bibinfo
  {volume} {153}},\ \bibinfo {pages} {275} (\bibinfo {year} {2003})},\ \Eprint
  {http://arxiv.org/abs/hep-ph/0301101} {arXiv:hep-ph/0301101} \BibitemShut
  {NoStop}%
\bibitem [{\citenamefont {Porod}\ and\ \citenamefont
  {Staub}(2012)}]{Porod:2011nf}%
  \BibitemOpen
  \bibfield  {author} {\bibinfo {author} {\bibfnamefont {W.}~\bibnamefont
  {Porod}}\ and\ \bibinfo {author} {\bibfnamefont {F.}~\bibnamefont {Staub}},\
  }\href {\doibase 10.1016/j.cpc.2012.05.021} {\bibfield  {journal} {\bibinfo
  {journal} {Comput. Phys. Commun.}\ }\textbf {\bibinfo {volume} {183}},\
  \bibinfo {pages} {2458} (\bibinfo {year} {2012})},\ \Eprint
  {http://arxiv.org/abs/1104.1573} {arXiv:1104.1573 [hep-ph]} \BibitemShut
  {NoStop}%
\bibitem [{\citenamefont {Staub}(2014)}]{Staub:2013tta}%
  \BibitemOpen
  \bibfield  {author} {\bibinfo {author} {\bibfnamefont {F.}~\bibnamefont
  {Staub}},\ }\href {\doibase 10.1016/j.cpc.2014.02.018} {\bibfield  {journal}
  {\bibinfo  {journal} {Comput. Phys. Commun.}\ }\textbf {\bibinfo {volume}
  {185}},\ \bibinfo {pages} {1773} (\bibinfo {year} {2014})},\ \Eprint
  {http://arxiv.org/abs/1309.7223} {arXiv:1309.7223 [hep-ph]} \BibitemShut
  {NoStop}%
\bibitem [{\citenamefont {Esteban}\ \emph {et~al.}(2019)\citenamefont
  {Esteban}, \citenamefont {Gonzalez-Garcia}, \citenamefont
  {Hernandez-Cabezudo}, \citenamefont {Maltoni},\ and\ \citenamefont
  {Schwetz}}]{Esteban:2018azc}%
  \BibitemOpen
  \bibfield  {author} {\bibinfo {author} {\bibfnamefont {I.}~\bibnamefont
  {Esteban}}, \bibinfo {author} {\bibfnamefont {M.}~\bibnamefont
  {Gonzalez-Garcia}}, \bibinfo {author} {\bibfnamefont {A.}~\bibnamefont
  {Hernandez-Cabezudo}}, \bibinfo {author} {\bibfnamefont {M.}~\bibnamefont
  {Maltoni}}, \ and\ \bibinfo {author} {\bibfnamefont {T.}~\bibnamefont
  {Schwetz}},\ }\href {\doibase 10.1007/JHEP01(2019)106} {\bibfield  {journal}
  {\bibinfo  {journal} {JHEP}\ }\textbf {\bibinfo {volume} {01}},\ \bibinfo
  {pages} {106} (\bibinfo {year} {2019})},\ \Eprint
  {http://arxiv.org/abs/1811.05487} {arXiv:1811.05487 [hep-ph]} \BibitemShut
  {NoStop}%
\bibitem [{\citenamefont {Aghanim}\ \emph {et~al.}(2020)\citenamefont {Aghanim}
  \emph {et~al.}}]{Aghanim:2018eyx}%
  \BibitemOpen
  \bibfield  {author} {\bibinfo {author} {\bibfnamefont {N.}~\bibnamefont
  {Aghanim}} \emph {et~al.} (\bibinfo {collaboration} {Planck}),\ }\href
  {\doibase 10.1051/0004-6361/201833910} {\bibfield  {journal} {\bibinfo
  {journal} {Astron. Astrophys.}\ }\textbf {\bibinfo {volume} {641}},\ \bibinfo
  {pages} {A6} (\bibinfo {year} {2020})},\ \Eprint
  {http://arxiv.org/abs/1807.06209} {arXiv:1807.06209 [astro-ph.CO]}
  \BibitemShut {NoStop}%
\bibitem [{\citenamefont {Baldini}\ \emph {et~al.}(2016)\citenamefont {Baldini}
  \emph {et~al.}}]{TheMEG:2016wtm}%
  \BibitemOpen
  \bibfield  {author} {\bibinfo {author} {\bibfnamefont {A.}~\bibnamefont
  {Baldini}} \emph {et~al.} (\bibinfo {collaboration} {MEG}),\ }\href {\doibase
  10.1140/epjc/s10052-016-4271-x} {\bibfield  {journal} {\bibinfo  {journal}
  {Eur. Phys. J. C}\ }\textbf {\bibinfo {volume} {76}},\ \bibinfo {pages} {434}
  (\bibinfo {year} {2016})},\ \Eprint {http://arxiv.org/abs/1605.05081}
  {arXiv:1605.05081 [hep-ex]} \BibitemShut {NoStop}%
\bibitem [{\citenamefont {Bellgardt}\ \emph {et~al.}(1988)\citenamefont
  {Bellgardt}, \citenamefont {Otter}, \citenamefont {Eichler}, \citenamefont
  {Felawka}, \citenamefont {Niebuhr}, \citenamefont {Walter}, \citenamefont
  {Bertl}, \citenamefont {Lordong}, \citenamefont {Martino}, \citenamefont
  {Egli}, \citenamefont {Engfer}, \citenamefont {Grab}, \citenamefont
  {Grossmann-Handschin}, \citenamefont {Hermes}, \citenamefont {Kraus},
  \citenamefont {Muheim}, \citenamefont {Pruys}, \citenamefont {{Van Der
  Schaaf}},\ and\ \citenamefont {Vermeulen}}]{BELLGARDT19881}%
  \BibitemOpen
  \bibfield  {author} {\bibinfo {author} {\bibfnamefont {U.}~\bibnamefont
  {Bellgardt}}, \bibinfo {author} {\bibfnamefont {G.}~\bibnamefont {Otter}},
  \bibinfo {author} {\bibfnamefont {R.}~\bibnamefont {Eichler}}, \bibinfo
  {author} {\bibfnamefont {L.}~\bibnamefont {Felawka}}, \bibinfo {author}
  {\bibfnamefont {C.}~\bibnamefont {Niebuhr}}, \bibinfo {author} {\bibfnamefont
  {H.}~\bibnamefont {Walter}}, \bibinfo {author} {\bibfnamefont
  {W.}~\bibnamefont {Bertl}}, \bibinfo {author} {\bibfnamefont
  {N.}~\bibnamefont {Lordong}}, \bibinfo {author} {\bibfnamefont
  {J.}~\bibnamefont {Martino}}, \bibinfo {author} {\bibfnamefont
  {S.}~\bibnamefont {Egli}}, \bibinfo {author} {\bibfnamefont {R.}~\bibnamefont
  {Engfer}}, \bibinfo {author} {\bibfnamefont {C.}~\bibnamefont {Grab}},
  \bibinfo {author} {\bibfnamefont {M.}~\bibnamefont {Grossmann-Handschin}},
  \bibinfo {author} {\bibfnamefont {E.}~\bibnamefont {Hermes}}, \bibinfo
  {author} {\bibfnamefont {N.}~\bibnamefont {Kraus}}, \bibinfo {author}
  {\bibfnamefont {F.}~\bibnamefont {Muheim}}, \bibinfo {author} {\bibfnamefont
  {H.}~\bibnamefont {Pruys}}, \bibinfo {author} {\bibfnamefont
  {A.}~\bibnamefont {{Van Der Schaaf}}}, \ and\ \bibinfo {author}
  {\bibfnamefont {D.}~\bibnamefont {Vermeulen}},\ }\href {\doibase
  https://doi.org/10.1016/0550-3213(88)90462-2} {\bibfield  {journal} {\bibinfo
   {journal} {Nuclear Physics B}\ }\textbf {\bibinfo {volume} {299}},\ \bibinfo
  {pages} {1 } (\bibinfo {year} {1988})}\BibitemShut {NoStop}%
\bibitem [{\citenamefont {Billard}\ \emph {et~al.}(2014)\citenamefont
  {Billard}, \citenamefont {Strigari},\ and\ \citenamefont
  {Figueroa-Feliciano}}]{Billard:2013qya}%
  \BibitemOpen
  \bibfield  {author} {\bibinfo {author} {\bibfnamefont {J.}~\bibnamefont
  {Billard}}, \bibinfo {author} {\bibfnamefont {L.}~\bibnamefont {Strigari}}, \
  and\ \bibinfo {author} {\bibfnamefont {E.}~\bibnamefont
  {Figueroa-Feliciano}},\ }\href {\doibase 10.1103/PhysRevD.89.023524}
  {\bibfield  {journal} {\bibinfo  {journal} {Phys. Rev. D}\ }\textbf {\bibinfo
  {volume} {89}},\ \bibinfo {pages} {023524} (\bibinfo {year} {2014})},\
  \Eprint {http://arxiv.org/abs/1307.5458} {arXiv:1307.5458 [hep-ph]}
  \BibitemShut {NoStop}%
\bibitem [{\citenamefont {Abbiendi}\ \emph {et~al.}(2003)\citenamefont
  {Abbiendi} \emph {et~al.}}]{Abbiendi:2003yd}%
  \BibitemOpen
  \bibfield  {author} {\bibinfo {author} {\bibfnamefont {G.}~\bibnamefont
  {Abbiendi}} \emph {et~al.} (\bibinfo {collaboration} {OPAL}),\ }\href
  {\doibase 10.1016/S0370-2693(03)00639-7} {\bibfield  {journal} {\bibinfo
  {journal} {Phys. Lett. B}\ }\textbf {\bibinfo {volume} {572}},\ \bibinfo
  {pages} {8} (\bibinfo {year} {2003})},\ \Eprint
  {http://arxiv.org/abs/hep-ex/0305031} {arXiv:hep-ex/0305031} \BibitemShut
  {NoStop}%
\bibitem [{\citenamefont {Ng}\ \emph {et~al.}(2017)\citenamefont {Ng},
  \citenamefont {Beacom}, \citenamefont {Peter},\ and\ \citenamefont
  {Rott}}]{Ng:2017aur}%
  \BibitemOpen
  \bibfield  {author} {\bibinfo {author} {\bibfnamefont {K.~C.~Y.}\
  \bibnamefont {Ng}}, \bibinfo {author} {\bibfnamefont {J.~F.}\ \bibnamefont
  {Beacom}}, \bibinfo {author} {\bibfnamefont {A.~H.~G.}\ \bibnamefont
  {Peter}}, \ and\ \bibinfo {author} {\bibfnamefont {C.}~\bibnamefont {Rott}},\
  }\href {\doibase 10.1103/PhysRevD.96.103006} {\bibfield  {journal} {\bibinfo
  {journal} {Phys. Rev. D}\ }\textbf {\bibinfo {volume} {96}},\ \bibinfo
  {pages} {103006} (\bibinfo {year} {2017})},\ \Eprint
  {http://arxiv.org/abs/1703.10280} {arXiv:1703.10280 [astro-ph.HE]}
  \BibitemShut {NoStop}%
\bibitem [{\citenamefont {Ferrer}\ \emph {et~al.}(2015)\citenamefont {Ferrer},
  \citenamefont {Ibarra},\ and\ \citenamefont {Wild}}]{Ferrer:2015bta}%
  \BibitemOpen
  \bibfield  {author} {\bibinfo {author} {\bibfnamefont {F.}~\bibnamefont
  {Ferrer}}, \bibinfo {author} {\bibfnamefont {A.}~\bibnamefont {Ibarra}}, \
  and\ \bibinfo {author} {\bibfnamefont {S.}~\bibnamefont {Wild}},\ }\href
  {\doibase 10.1088/1475-7516/2015/09/052} {\bibfield  {journal} {\bibinfo
  {journal} {JCAP}\ }\textbf {\bibinfo {volume} {09}},\ \bibinfo {pages} {052}
  (\bibinfo {year} {2015})},\ \Eprint {http://arxiv.org/abs/1506.03386}
  {arXiv:1506.03386 [hep-ph]} \BibitemShut {NoStop}%
\bibitem [{\citenamefont {Aartsen}\ \emph {et~al.}(2020)\citenamefont {Aartsen}
  \emph {et~al.}}]{Amole:2019coq}%
  \BibitemOpen
  \bibfield  {author} {\bibinfo {author} {\bibfnamefont {M.~G.}\ \bibnamefont
  {Aartsen}} \emph {et~al.} (\bibinfo {collaboration} {IceCube, PICO}),\ }\href
  {\doibase 10.1140/epjc/s10052-020-8069-5} {\bibfield  {journal} {\bibinfo
  {journal} {Eur. Phys. J. C}\ }\textbf {\bibinfo {volume} {80}},\ \bibinfo
  {pages} {819} (\bibinfo {year} {2020})},\ \Eprint
  {http://arxiv.org/abs/1907.12509} {arXiv:1907.12509 [astro-ph.HE]}
  \BibitemShut {NoStop}%
\bibitem [{\citenamefont {Steigman}\ \emph {et~al.}(2012)\citenamefont
  {Steigman}, \citenamefont {Dasgupta},\ and\ \citenamefont
  {Beacom}}]{Steigman:2012nb}%
  \BibitemOpen
  \bibfield  {author} {\bibinfo {author} {\bibfnamefont {G.}~\bibnamefont
  {Steigman}}, \bibinfo {author} {\bibfnamefont {B.}~\bibnamefont {Dasgupta}},
  \ and\ \bibinfo {author} {\bibfnamefont {J.~F.}\ \bibnamefont {Beacom}},\
  }\href {\doibase 10.1103/PhysRevD.86.023506} {\bibfield  {journal} {\bibinfo
  {journal} {Phys. Rev. D}\ }\textbf {\bibinfo {volume} {86}},\ \bibinfo
  {pages} {023506} (\bibinfo {year} {2012})},\ \Eprint
  {http://arxiv.org/abs/1204.3622} {arXiv:1204.3622 [hep-ph]} \BibitemShut
  {NoStop}%
\bibitem [{\citenamefont {Navarro}\ \emph {et~al.}(1996)\citenamefont
  {Navarro}, \citenamefont {Frenk},\ and\ \citenamefont
  {White}}]{Navarro:1995iw}%
  \BibitemOpen
  \bibfield  {author} {\bibinfo {author} {\bibfnamefont {J.~F.}\ \bibnamefont
  {Navarro}}, \bibinfo {author} {\bibfnamefont {C.~S.}\ \bibnamefont {Frenk}},
  \ and\ \bibinfo {author} {\bibfnamefont {S.~D.~M.}\ \bibnamefont {White}},\
  }\href {\doibase 10.1086/177173} {\bibfield  {journal} {\bibinfo  {journal}
  {Astrophys. J.}\ }\textbf {\bibinfo {volume} {462}},\ \bibinfo {pages} {563}
  (\bibinfo {year} {1996})},\ \Eprint {http://arxiv.org/abs/astro-ph/9508025}
  {arXiv:astro-ph/9508025} \BibitemShut {NoStop}%
\end{thebibliography}%

\end{document}